\documentclass[preprint,12pt]{elsarticle}
\journal{Information Systems}

\usepackage{amssymb}
\usepackage{amsmath}
\usepackage{algorithm}
\usepackage{algorithmicx}
\usepackage{algpseudocode}
\usepackage[utf8x]{inputenc}
\usepackage{graphicx}
\usepackage{subfigure}
\usepackage{multicol}
\usepackage{multirow}
\usepackage{xspace}
\usepackage[usenames,dvipsnames]{xcolor}

\mathchardef\dash="2D 

\renewcommand{\k}{\ensuremath{k}\xspace}
\newcommand{\kk}{k\ensuremath{^2}\xspace}
\newcommand{\ktree}{k\ensuremath{^2}-tree\xspace}
\newcommand{\ktrees}{k\ensuremath{^2}-trees\xspace}
\newcommand{\ktriples}{k\ensuremath{^2}-triples\xspace}
\newcommand{\ktriplesplus}{k\ensuremath{^2}-triples$^+$\xspace}
\newcommand{\dktree}{dk\ensuremath{^2}-tree\xspace}
\newcommand{\dktrees}{dk\ensuremath{^2}-trees\xspace}
\newcommand{\ttree}{T\ensuremath{_{tree}}\xspace}
\newcommand{\ltree}{L\ensuremath{_{tree}}\xspace}

\newcommand{\irank}{\emph{rank}\xspace}
\newcommand{\iselect}{\emph{select}\xspace}
\newcommand{\iaccess}{\emph{access}\xspace}
\newcommand{\rank}{rank\xspace}
\newcommand{\select}{select\xspace}
\newcommand{\access}{access\xspace}

\begin{document}
\begin{frontmatter}
\title{Compressed Representation of Dynamic Binary Relations with Applications\tnoteref{t1}}
\tnotetext[t1]{\footnotesize Funded in part by European Union's Horizon 2020 research and innovation programme
     under the Marie Sklodowska-Curie grant agreement No 690941 (project BIRDS).
  G. Navarro was partially funded by a Google Research Award Latin America and
  by Fondecyt [1-140796].
  N. Brisaboa, A. Cerdeira-Pena, and G. de Bernardo were partially funded
    by  MINECO (PGE, CDTI, and FEDER) [TIN2013-46238-C4-3-R, IDI-20141259,
    ITC-20151305, ITC-20151247, TIN2015-69951-R, ITC-20161074,
    TIN2016-78011-C4-1-R] by ICT COST Action IC1302; and by Xunta de Galicia
    (co-funded with ERDF) [GRC2013/053, Centro singular de investigaci\'on de
    Galicia accreditation 2016-2019].
  An early partial version of this article appeared in {\em Proc DCC'12}  \cite{BDNDCC12}.}

    \author[udc]{Nieves R. Brisaboa}
    \ead{brisaboa@udc.es}

    \author[udc]{Ana Cerdeira-Pena}
    \ead{acerdeira@udc.es}

    \author[udc]{Guillermo de Bernardo\corref{cor1}}
    \ead{gdebernardo@udc.es}

    \author[uchile]{Gonzalo Navarro}
    \ead{gnavarro@dcc.uchile.cl}

    \cortext[cor1]{Corresponding author}

    \address[udc]{Database Laboratory, University of A Coru\~na, Spain.\\}    
    \address[uchile]{Department of Computer Science, University of Chile, Chile.}


\begin{abstract}
We introduce a dynamic data structure for the compact representation
of binary relations $\mathcal{R} \subseteq A \times B$. The data
structure is a dynamic variant of the \ktree, a static compact
representation that takes advantage of clustering in the binary
relation to achieve compression. Our structure can efficiently check
whether two objects $(a,b) \in A \times B$ are related, and list the
objects of $B$ related to some $a \in A$ and vice versa.
Additionally, our structure allows inserting and deleting pairs
$(a,b)$ in the relation, as well as modifying the base sets $A$ and
$B$. We test our dynamic data structure in different contexts,
including the representation of Web graphs and RDF databases. Our
experiments show that our dynamic data structure achieves good
compression ratios and fast query times, close to those of a static
representation, while also providing efficient support for updates
in the represented binary relation.
\end{abstract}


\end{frontmatter}

\section{Introduction} \label{section:Introduction}
Binary relations arise everywhere in Computer Science: graphs,
matchings, discrete grids or inverted indexes in document retrieval are just some examples. Consider a binary relation between
two sets $A$ and $B$, defined as a subset $\mathcal{R} \subseteq A \times B$.
Typical operations of interest in a binary relation are: determine whether a pair $(a,b)$
is in $\mathcal{R}$, find all the elements $b\in B$ such that
$(a,b)\in \mathcal{R}$, given $a\in A$, and vice versa. More
sophisticated ones aim, for example, at retrieving all pairs $(a,b)
\in \mathcal{R}$ where $a \in [a_1,a_2]$ and $b \in [b_1,b_2]$.

Web graphs, where nodes are Web pages and relations are hyperlinks,
can be seen as a binary relation between two (usually equal) sets of
Web pages $A$ and $B$. In this context, basic binary relation
operations are translated into queries to find the direct or reverse
neighbors of a node. The ``range'' query involving all pairs $(a,b)
\in \mathcal{R}$ where $a \in [a_1,a_2]$ and $b \in [b_1,b_2]$ can
be used to retrieve all the links between two Web sites, considering
that Web pages are sorted lexicographically and therefore all pages
in a Web site are consecutive in an ordering of the base sets. In
the context of document retrieval, an inverted index can be seen as
a binary relation between a set of documents $D$ and a set of terms
(usually words) $\Sigma$. In this context, we can use binary
relation operations to find all the documents where a term $w$
appears (all $d \in D$ where $(w,d) \in \mathcal{R} \subseteq D
\times \Sigma$) or to find whether a term appears in a document
(checking if $(w,d) \in \mathcal{R}$).

In addition to the previous examples, other multidimensional data
can be naturally represented as collections of binary relations. A
usual case occurs when a dataset contains ``labeled'' relations
between two base sets (that is, the relation itself has a property
or label); in this case, a 3-dimensional dataset can actually be
seen as a collection of binary relations, one for each value in the
third dimension. A good example of this kind of datasets are RDF
(Resource Description Framework) graphs. RDF is a standard for the
representation of knowledge in the Web of Data. RDF graphs are
labeled graphs with a set of subject (origin) nodes $S$, a set of
target (object) nodes and a set of predicates (labels) $P$. An edge
in an RDF graph represents a property of element $s$, given by
predicate $p$ and with value $o$. A usual strategy to store and
query these datasets is to apply a vertical partitioning strategy
\cite{abadi2007} to divide the data by predicate, since the number
of predicates is generally small. Through vertical partitioning, an
RDF graph can be transformed into a collection of binary relations
$\mathcal{R}_p \subseteq (S,O)$ for each $p \in P$, representing the
valid pairs $(s,o)$ for each predicate.

There are two natural ways to represent binary relations: a binary
adjacency matrix or an adjacency list. On large binary relations,
reducing space while retaining functionality is crucial in order to
operate efficiently in main memory. Therefore, simple
representations such as plain adjacency matrices are usually
unfeasible in these datasets. On the other hand, simple adjacency
lists can efficiently compress sparse binary relations, but an
adjacency list representation usually lacks the ability to answer
queries symmetrically, or to efficiently retrieve information on
ranges of elements. The limitations of simple data structures has
led to different proposals for compressing general binary
relations~\cite{BCN10}, as well as specific ones such as Web
graphs~\cite{Boldi03}.

Brisaboa et al.~\cite{Brisaboa09} introduced a compact data
structure called {\em \ktree}. It was initially proposed for the
compression of Web graphs, where it was shown to be very competitive
(see also \cite{Claude11}). Since then, it has also been
successfully applied to other domains such as RDF databases
\cite{Alvarez11} and social networks \cite{Claude11}. In fact,
\ktrees can be used for the representation of general binary
relations and take advantage of clustering in the binary matrix to
achieve compression. They support elegantly all the described
operations (simple and sophisticated) as instances of the most
general range query.

However, just like the other compressed representations of graphs
and binary relations, \ktrees are essentially static. This
discourages their use in cases where the binary relation changes due
to the insertion or deletion of pairs $(a,b)$ (e.g., adding or
removing edges in a graph) or of elements in $A$ and $B$ (e.g.,
adding or removing graph nodes, or words or documents in inverted
indexes).

Dynamic representations of compact data structures are usually affected by a
slowdown factor over the equivalent static data structure~\cite[Chapter
12]{Nav16}.
For example, a dynamic bitmap has a lower bound of $\Omega(\frac{\log n}{\log \log n})$ for
many operations that static bitmaps solve in $\mathcal{O}(1)$. Another example
exists in 2-dimensional grids, that are queried by \ktrees: range reporting
queries have a complexity of $\mathcal{O}(\frac{\log n}{\log \log n})$ in a
static representation, but a lower bound of $\Omega((\frac{\log
n}{\log \log n})^2)$ exists in a dynamic approach. Hence dynamic representations
of a structure like the \ktree are expected to be slower, and larger, 
than a static representation, in many cases. In practice, in applications
where update operations are required, the slowdown factor of the dynamic
representation and the frequency of updates become key to determine which is
the best approach.

In this paper we introduce the {\em \dktree}, a dynamic version of
the \ktree. Our data structure achieves space
utilization close to that of the static structure, and allows the insertion and deletion of
pairs and elements in the sets (i.e., changing bits and
inserting/deleting rows/columns in the binary matrix). Our
experiments show that \dktrees achieve good space/time tradeoffs in
comparison with the equivalent static representation. In Web graphs,
where \ktrees obtained good compression results and query times, our
dynamic representation obtains query times less than twice those of
the static representation. Our results also show that, depending on
the characteristics of the datasets, update operations in the
\dktree can also be as efficient as queries.

We apply our proposal to the representation of RDF databases, where
static \ktrees were competitive with state-of-the-art alternatives
but lacked the update capabilities usually required in this kind of
graphs \cite{kafer2013observing}. We show that our representation
can easily answer all queries supported by static \ktrees using
similar algorithms, while providing update capabilities. Our dynamic
data structure only requires a 20-50\% space overhead to store the
dataset, and requires less than twice the query times of the
equivalent static representation to answer most of the queries.
We choose RDF databases as an example where static \ktrees have already
been used but are limited by their static nature. However, there are several other
application domains where the use of a dynamic data structure for the compact representation
of binary relations could also be worthwhile: time-evolving regions
(e.g. oil stains), communication networks, social graphs, etc.

\section{Related Work} \label{section:RelatedWork}

\subsection{Previous Concepts}
In this section we present some necessary background to better
understand our contribution and to make the manuscript
self-contained.

\subsubsection{Rank and select over bitmaps} \label{sec:previous-concepts:rank}

Bit vectors (often referred to as bitmaps, bit strings, etc.)
supporting \irank and \iselect operations~\cite{Jac89} are the basis
of many other succinct data structures. We next describe them in
detail.

Let be $B[1,n]$ a binary sequence of size $n$. Then \irank and
\iselect are defined as:

\begin{itemize}
\item $\rank_b(B,p) = i$ if the number of occurrences of the bit $b$ from the beginning of $B$ up to position $p$ is $i$.
\item $\select_b(B,i) = p$ if the $i \dash th$ occurrence of the bit $b$ in the
sequence $B$ is at position $p$.
\end{itemize}

Given the importance of these two operations in the performance of
other succinct data structures, like \emph{full-text
indexes}~\cite{NM07}, many strategies have been developed to
efficiently implement \irank and \iselect.

Jacobson~\cite{Jac89} proposed an implementation for this problem
able to compute \irank in constant time. It is based on a two-level
directory structure. The first-level directory stores $\rank_b(B,p)$
for every $p$ multiple of $s=\left \lfloor \log n \right \rfloor
\left \lfloor \log{n/2} \right \rfloor$. The second-level directory
holds, for every $p$ multiple of $b=\left \lfloor \log {n/2} \right
\rfloor$, the relative \irank value from the previous multiple of
$s$. Following this approach, $\rank_1(B,p)$ can be computed in
constant time adding values from both directories: the first-level
directory returns the \irank value until the previous multiple of
$s$. The second-level directory returns the number of ones until the
previous multiple of $b$. Finally, the number of ones from the
previous multiple of $b$ until $p$ is computed sequentially over the
bit vector. This computation can be performed in constant time using
a precomputed table that stores the \irank values for all possible
block of size $b$. As a result, \irank can be computed in constant
time. The \iselect operation can be solved using binary searches in
$O(\log \log n)$ time. The sizes $s$ and $b$ are carefully chosen so
that the overall space required by the auxiliary dictionary
structures is $o(n)$: $O(n/\log n)$ for the first-level directory,
$O(n \log \log n / \log n)$ for the second-level directory and
$O(\log n \log \log n)$ for the lookup table. Later works by
Clark~\cite{Cla96} and Munro~\cite{Mun96} obtained constant time
complexity also for the \iselect operation, using additional $o(n)$
space. For instance, Clark proposed a new three-level directory
structure that solved $select_1$, and could be duplicated to also
answer $select_0$.

The previous proposals solve the problem, at least in theory, of
adding \irank and \iselect support over a bit vector using $o(n)$
additional bits. However, further work was devoted to obtain even
more compressed representations, taking into account the actual
properties of the binary sequence~\cite{Pag99, RRR02, OS07}.



Another alternative study, called \emph{gap encoding}, aims to
compress the binary sequences when the number of $1$ bits is small.
It is based on encoding the distances between consecutive $1$ bits.
Several developments following this approach have been
presented~\cite{Sad03, GGV04, BB04, GHSV06, MN07}.

\subsubsection{ETDC and DETDC} \label{sec:previous-concepts:etdc}

\paragraph*{End-Tagged Dense Code (ETDC)} It is a
semi-static statistical byte-oriented encoder/decoder  \cite{etdc,
etdcj}, that achieves very good compression and decompression times
while keeping similar compression ratios to those obtained by Plain
Huffman~\cite{huffman} (the byte-oriented version of Huffman
\cite{H52} that obtains optimum byte-oriented prefix codes).

Consider a sequence of symbols $D=d_1 \ldots d_n$. In a first pass
ETDC computes the frequency of each different symbol in the
sequence, and creates a vocabulary where the symbols are placed
according to their overall frequency in descending order. ETDC
assigns to each entry of the vocabulary a variable-length code, that
will be shorter for the first entries of the vocabulary (more
frequent symbols). Then, in a second pass, each symbol of the
original sequence is replaced by the corresponding variable-length
code.

The key idea in ETDC is to mark the end of each codeword
(variable-length code): the first bit of each byte will be a flag,
set to 1 if the current byte is the last byte of a codeword, or 0
otherwise. The remaining 7 bits in the byte are used to assign the
different values sequentially, which makes the codeword assignment
extremely simple in ETDC. Consider the symbols of the vocabulary,
that are stored in descending order by frequency: the first 128
($2^7$) symbols will be assigned 1-byte codewords, the next $128^2$
symbols will be assigned 2-byte codewords, and so on. The codewords
are assigned depending only on the position of the symbol in the
sorted vocabulary. The simplicity of the code assignment is the
basis for the fast compression and decompression times of ETDC. In
addition, its ability to use all the possible combinations of 7 bits
to assign codewords makes it very efficient in space. Notice also
that ETDC can work with a different chunk size for the codewords: in
general, we can use any chunk of size $b$, using 1 bit as flag and
the remaining $b-1$ bits to assign codes, hence having $2^{b-1}$
codewords of 1 chunk, $2^{2(b-1)}$ codewords of 2 chunks and so on.
Nevertheless, bytes are used as the basic chunk size ($b=8$) in most
cases for efficiency.

\paragraph*{Dynamic End-Tagged Dense Code (DETDC)} It is an adaptive
(one-pass) version of ETDC \cite{detdc}. 
As an adaptive mechanism, it does not require to preprocess and sort
all the symbols in the sequence before compression. Instead, it
maintains a vocabulary of symbols that is modified according to the
new symbols received by the compressor.

The solution of DETDC for maintaining an adaptive vocabulary is to
keep the vocabulary of symbols always sorted by frequency. This
means that new symbols are always appended at the end of the
vocabulary (with frequency 1), and existing symbols may change their
position in the vocabulary when their frequency changes during
compression.

The process for encoding a message starts by reading the message
sequentially. Each symbol read is looked up in the vocabulary, and
processed depending on whether it is found or not:
\begin{itemize}
  \item If the symbol is not found in the vocabulary it is a new symbol,
  therefore it is appended at the end of the vocabulary with frequency 1.
  The encoder writes the new codeword to the output, followed by the
  symbol itself. The decoder can identify a new symbol because its
  codeword is larger than the decoder's vocabulary size, and add the
  new symbol to its own vocabulary.
  \item If the symbol is found in the vocabulary, the encoder simply
  writes its codeword to the output. After writing to the output, the
  encoder updates the frequency of the symbol (increasing it by 1) and
  reorders the vocabulary if necessary. Since the symbol frequency has
  changed from $f$ to $f+1$, it is moved to the region where symbols with
  frequency $f+1$ are stored in the vocabulary. This reordering process
  is performed swapping elements in the vocabulary. The key of DETDC is
  that the encoder and the decoder share the same model for the vocabulary
  and update their vocabulary in the same way, so changes in the vocabulary
  during compression can be automatically performed by the decoder using the
  same algorithms without transmitting additional information.
\end{itemize}

DETDC and its variants are able to obtain very good results to
compress natural language texts, obtaining compression very close to
original ETDC without the first pass required by the semi-static
approach.


\subsection{The \ktree}
\label{sec:ktree}

\begin{figure*}[ht]
\begin{center}
   \includegraphics[width=\textwidth]{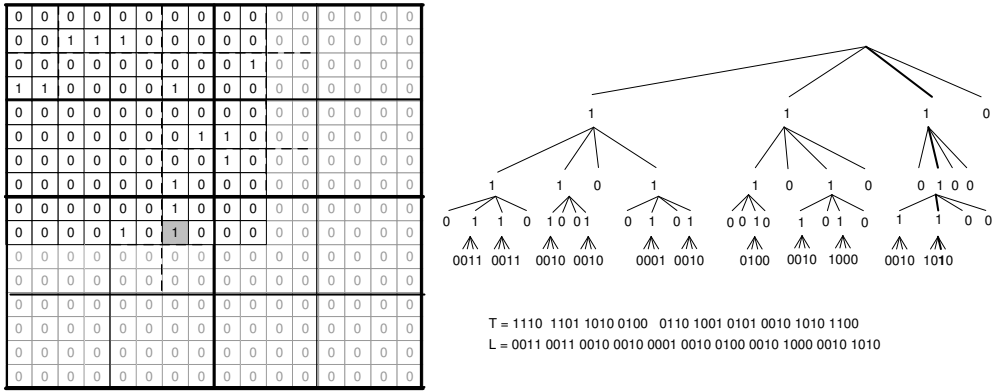}
   \caption{Example of \ktree for a binary matrix using $\k=2$.}
   \label{fig:k2tree}
\end{center}
\end{figure*}

A \ktree is conceptually a \kk-ary tree built by recursively
partitioning a binary matrix. At each partitioning step, the current
matrix of size $n \times n$ is divided into \kk submatrices of size
$n/k \times n/k$ \footnote{The size of the matrix is assumed to be a
power of \k. If $n$ is not a power of \k, we use instead $n' =
\k^{\lceil \log n \rceil}$, the next power of \k. Conceptually, the
matrix is expanded with new zero-filled rows and columns until it
reaches $n' \times n'$.}. Figure~\ref{fig:k2tree} shows a
$10\times10$ binary matrix, virtually expanded to size $16\times16$,
and the conceptual \ktree that represents it, for $\k=2$. The
submatrices are numbered from 0 to \kk-1, starting from left to
right and top to bottom. The first level of the tree contains one
node with \kk children, representing the \kk submatrices in which
the original matrix is divided following a quadtree-like
subdivision. Each node is represented using a single bit: 1 if the
submatrix has at least one cell with value 1, or 0 otherwise. A 0
child means that there are no ones in the corresponding submatrix,
and it has no children. The decomposition continues recursively for
each 1 child until the current submatrix is full of zeros or we
reach the individual cells of the original matrix. The underlying
conceptual tree is in fact an MX-Quadtree~\cite{Samet84}, that
recursively decomposes the space in four quadrants stopping only 
when the region is fully empty (all cells set to 0) or when the maximum
precision is reached (individual cells of the adjacency matrix). Hence, a \ktree
that uses $\k=2$ can be seen as a compact and efficient representation of this conceptual quadtree.

This conceptual tree is implemented using two bit arrays: $T$
contains the bits for all the levels of the tree except for the last
one, taken in a levelwise traversal of the tree. $L$ stores the bits
of the last level of the tree.

The \ktree allows navigation of the conceptual tree using only the
bitmap representations thanks to a basic property: given any
internal node in the \ktree (a position $pos$ in $T$), its \kk
children will be located at $pos'=rank_1(T, pos) \times \kk$,
because each bit set to one adds \kk bits to the next level and bits
set to zero do not have descendants. If the position exceeds the
length of $T$, $L[pos'-|T|]$ is used. A rank structure is built over
T to provide an efficient $rank_1$ operation. All query operations
are based on this basic navigation of the conceptual tree.

To access a cell of the matrix, the tree is navigated from the root
until a 0 is found or the last level is reached. At each level, the
child whose submatrix contains the target cell is selected. If the
bit value of that node is 0 we know the cell in the region is a 0,
and navigation ends. If the value is 1, we proceed recursively to
the appropriate children. For instance, let us suppose we want to
retrieve the value of the cell at row 9, column 6 in the matrix of
Figure~\ref{fig:k2tree}. The path to reach that cell has been
highlighted in the conceptual \ktree. To perform this
navigation\footnote{We refer the reader to \cite{BLN14} for specific
implementation details.}, we would start at the root of the tree
(position 0 in $T$). In the first level, we need to access the third
child (offset 2), since we are accessing the bottom-left quadrant; hence we
access position $2$ in $T$.
Since $T[2]=1$, we know we are in an internal node. Its children will
begin at position $\rank_1(T,2) \times \kk = 12$, where we find the
bits \texttt{0100}. In this level we must access the second child
(offset 1), so we check $T[12+1]=T[13]=1$. Again, we are at an
internal node, and its children are located at position $\rank_1(T,
13) \times \kk = 9 \times 4 = 36$.  We have reached the third level
of the conceptual tree, and we need to access now the second child
(offset 1). Again, $T[36+1] = 1$, so we compute the position of its
children using $p = \rank_1(T,36) \times 4 = 80$. Now $p$ is higher
than the size of $T$ (40), so the \kk bits will be located at
position $p-|T|=80-40=40$ in $L$. Finally, in $L$ we find the \kk
bits \texttt{1010}, and need to check the third element. We find a 1
in $L$ and return the result.

In addition to the retrieval of a single cell of the matrix, \ktrees
can perform other operations efficiently: with some additional
calculations, we can modify the basic search to find all the ones in
a row/column, or a range [a$_1$,a$_2$]-[b$_1$,b$_2$]. To do this, at
each level of the \ktree we must access all the children of the
current node that overlap the region we are interested in,
traversing all the branches of the tree that intersect the region of
interest.  The cost to perform a general range reporting query 
in a \ktree, over a region of size $p \times q$, is bounded by the size of
the region and the number $occ$ of results found. The upper bound for the query
time is $O(p+q+(occ+1)k \log_k s)$, where $s$ is the size of the
longest side of the matrix~\cite[Section 10.2.1]{Nav16}.

\subsubsection{Improvements}
\label{sec:k2tree:improvements}

Several enhancements have been proposed to obtain better compression
results in the \ktree (see~\cite{Ladra11}). The first modification
is the use of different values of $k$ in different levels of the
\ktree. By using a bigger $k$ for the first levels and a smaller
value for the remaining ones, one can achieve better query times
(because the \ktree's height is reduced) with good space results.
This is called a \emph{hybrid} \ktree representation.

Another major improvement proposed over basic \ktrees is the use of
a compression method for the bitmap $L$. In this approach, the
lowest levels of the \ktree are grouped, yielding submatrices of
size $k' \times k'$ (for instance, to use $8 \times 8$ submatrices
as the last level of the tree instead of $2 \times 2$). These
submatrices are sorted according to their overall frequency and
stored in a matrix vocabulary $L^{voc}$. Then, the bitmap $L$ is
replaced by a sequence of variable-length codes $L^{seq}$. The
variable-length codes are assigned using (s,c)-Dense
Codes~\cite{scdc}, and $L^{seq}$ is encoded using Direct Access
Codes~\cite{Brisaboa09DAC} to provide direct access to any entry in
the sequence. This variant can reduce significantly the size of the
representation while showing similar query times.

\section{The dynamic \ktree: \dktree} \label{sec:kdyn:structure}

\subsection{Data structures}
\label{sec:kdyn:structure:structures}

The conceptual \ktree is represented, in its static variant, using
two bit arrays, $T$ and $L$. In our dynamic version, we represent
$T$ and $L$ with two trees, that we call \ttree and \ltree. Our
approach to represent the \ktree using these trees is called
\emph{\dktree}. Our trees, \ttree and \ltree, are in fact practical
implementations of dynamic bit vectors~\cite{Raman01} replacing the
static bitmaps $T$ and $L$. The leaves of \ttree and \ltree contain
roughly the bits in $T$ and $L$, while the internal nodes provide
access to arbitrary positions and also act as a dynamic rank
structure.

\begin{figure*}[ht]
\begin{center}
   \includegraphics[width=0.8\textwidth]{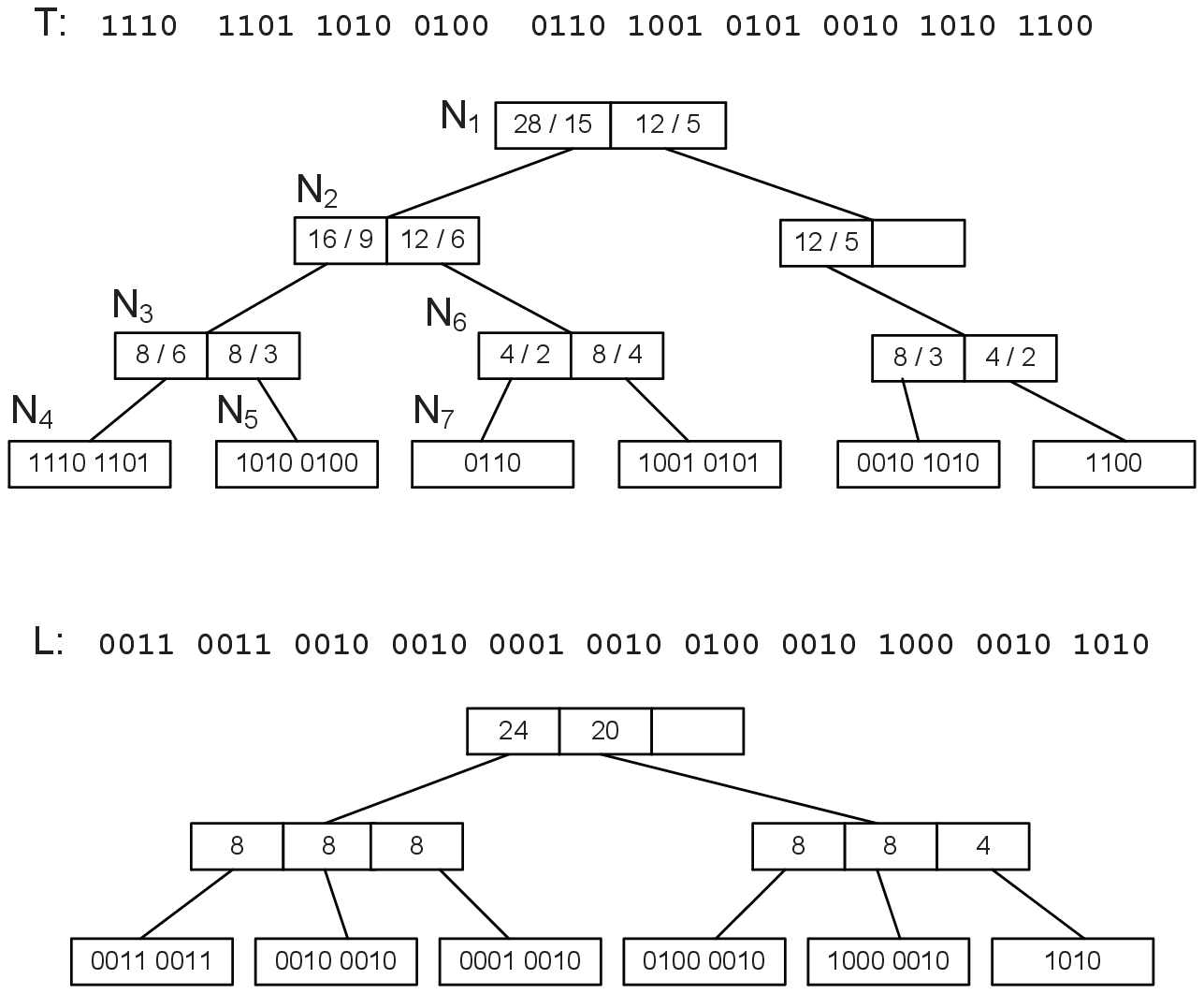}
   \caption{Dynamic \ktree representation.}
   \label{fig:dk2tree}
\end{center}
\end{figure*}

Consider a static \ktree representation (bitmaps $T$ and $L$). To
build a \dktree representation from it, we partition $T$ and $L$ in
blocks of up to $B$ bits. The generated blocks will be the leaves of
\ttree and \ltree. The internal nodes of our trees contain a set of
entries that allow us to access the leaves for query and update
operations. Each entry in \ttree is of the form $\langle b,o,P
\rangle$, where $b$ and $o$ are counters and $P$ is a pointer to the
corresponding child node. The values $b$ and $o$ in each entry will
allow us to efficiently access and perform \irank operations in the
dynamic bitmaps. If $P$ points to a leaf node, the counters will
store the number $b$ of bits stored in the leaf and the number of
them that are \emph{ones} ($o$). If $P$ points to an internal node,
$b$ and $o$ will contain the sum of all the $b\dash$ and
$o\dash$counters in the child node. Internal nodes in \ltree are
very similar, but entries only store values $\langle b, P \rangle$,
since \irank support in $L$ is not needed. Figure~\ref{fig:dk2tree}
shows a \dktree representation for the \ktree of
Figure~\ref{fig:k2tree}. Values of $b\dash$ and $o\dash$counters are
represented in the nodes, and pointers are visually represented.

The nodes of \ttree and \ltree may in general be partially empty.
Each node has a maximum and minimum capacity and may contain any
number of bits or entries between those parameters. A $size$ field
is added to keep the current size of a node. In leaf nodes, this
field contains the number of bits stored in the block. In internal
nodes, it stores the number of fixed-size entries used. The tree is
completely balanced, and nodes may be split or merged when the
contents change. The behavior of \ttree and \ltree on update
operations will be explained in more detail in Section
\ref{sec:kdyn:structure:update}.

\subsection{Query algorithms}
\label{sec:kdyn:structure:query}

All the queries supported by \ktrees are based on \iaccess and
\irank operations over the bitmaps $T$ and $L$. As explained before,
our tree structures, \ttree and \ltree, are essentially dynamic
replacements for the static bitmaps. By providing basic support for
these two simple operations in the bitmaps stored by \ttree and
\ltree, all the queries supported by static \ktrees can be directly
supported by our \dktree.

\newcommand{\findleaf}{\emph{findLeaf}\xspace}
\newcommand{\bbefore}{\ensuremath{b_{\mathrm{before}}}\xspace}
\newcommand{\obefore}{\ensuremath{o_{\mathrm{before}}}\xspace}
\newcommand{\size}{\mathit{size}}
\newcommand{\leaf}{\mathit{leaf}}
\newcommand{\nentries}{\mathit{nEntries}}

The navigation of a conceptual \ktree is based on a sequence of
\iaccess operations, to check the value of a node, followed by
possible \irank operations to locate its children. The \iaccess
operation, that is trivial in a static bitmap, is decomposed in our
\ttree and \ltree in two steps: first, an operation \findleaf is
used to locate the leaf that contains the desired bitmap, and the
offset in the leaf node where the bit should be; then, we access the
leaf node's bitmap to retrieve the actual values. This \findleaf
operation also computes information about the number of ones up to
the beginning of the leaf node, to allow efficient computation of
\irank operations if necessary.

Two slightly different algorithms are used to access a leaf in
\ttree and \ltree, but the essential steps are the same. Starting at
the root of \ttree or \ltree, the entries of the current node are
checked from left to right, accumulating the values of $b\dash$ and
$o\dash$counters in two variables, \bbefore and \obefore, until we
exceed the desired position $p$. The algorithms proceed to the
corresponding child node that would contain $p$. When a leaf node is
found, the values of \bbefore and \obefore contain the bit count and
\irank to the left of the node.

Once \findleaf returns the leaf node $N_\ell$ that contains the
desired position, we can \iaccess the desired bit at position $p -
\bbefore$ in the bitmap of the leaf node ($T_\ell$). If the bit has
value 0, navigation ends. Otherwise, and assuming we are still in
the upper levels of the conceptual \ktree, we would need to locate
its children, that will be located at position $\irank_1(\ttree,
p)\times \kk$. The rank value is computed as $\obefore +
\rank_1(T_\ell, p - \bbefore)$.

The \irank operation in $T_\ell$ can still be a costly operation for
relatively large node sizes. In order to speed up the local rank
operation inside the leaves of \ttree (\emph{rankLeaf} operation),
we add a small rank structure to each leaf of \ttree. This rank
structure is simply a set of counters that stores the number of ones
in each block of $S_T$ bits. $S_T$ determines the number of samples
that are stored in each leaf and provides a space/time tradeoff for
the local rank operation inside the leaves of \ttree. Using this
modified leaf structure, $\rank_1(T_\ell, p)$ is obtained adding the
values of all samples previous to position $p$ and performing a
sequential counting operation only from the last sampled
position\footnote{With a cost comparable to that of many practical
static rank structures.}.

\subsubsection{Improving access times}
\label{improving}

An actual query in a \ktree involves usually a top-down traversal,
following a number of branches of the conceptual tree. This top-down
traversal actually translates into a set of accesses to the bitmaps
$T$ and $L$. These accesses follow a well-defined pattern, starting
at the beginning of the bitmap and accessing new positions left to
right.

\newcommand\findleafstar{\ensuremath{\mathit{findLeaf}^*}\xspace}

Taking advantage of this property, we propose an alternative
strategy to navigate the \dktree. In this strategy, to \iaccess a
position in \ttree or \ltree we start the search from the previously
accessed leaf instead of the root node. Instead of a fixed number of
internal nodes to be traversed top-down, the new algorithm will
first traverse the tree bottom-up until a descendant of the new node
is found, and then continue top-down as the original algorithm. This
method aims at taking advantage of the access patterns in the \ktree
bitmaps, since many accesses, especially in upper levels, will be
located in the same or very close leaf nodes in \ttree.

To be able to start search from a previously accessed leaf node, a
new \findleafstar operation must store a small array containing
information about the last traversed path.
\emph{levelData}$[$\emph{\ttree.depth}$]$ is kept, that stores for
each level of \ttree, an entry $\langle N, e, s, b, o \rangle$,
where $N$ is the \ttree node accessed at that level, $e$ the entry
that was traversed, $s$ is the number of bits covered by $N$ and $b$
and $o$ are the values of \bbefore and \obefore. A similar array is
kept for \ltree, only ignoring the $o$ values in each entry.

The path information from the previous call allows \findleafstar to
determine whether the current leaf node contains the new desired
position. If it does, the method returns immediately. Otherwise, the
parent node is checked recursively until we find an internal node
that ``covers'' the new position. From that point on, the algorithm
behaves exactly like the original \findleaf and its top-down
traversal.

\subsection{Update operations}
\label{sec:kdyn:structure:update}

In addition to the queries supported by static \ktrees, \dktrees
must support update operations over the binary relation. First, relations
between existing elements may be created or deleted (changing zeros
of the adjacency matrix into ones and vice versa). Additionally,
\dktrees support changes in the base sets of the binary relation
(new rows/columns can be added to the binary adjacency matrix, and existing
rows/columns can be removed as well).

\subsubsection{Changing the contents of the binary matrix}

Changes in the binary matrix represented by a \ktree lead to a set
of modifications in the conceptual tree representation, essentially
the creation or removal of branches in this conceptual \ktree. We
will describe the changes caused in the conceptual tree and its
bitmap representation. Then we will explain how these changes in the
bitmaps are implemented over the data structures \ttree and \ltree.

In order to insert a new 1 in a binary matrix represented with a
\ktree, we need to make sure that an appropriate path exists in the
conceptual tree that reaches the cell. The insertion procedure
begins searching for the cell that has to be inserted, until a 0 is found in the
conceptual tree. Two cases may occur:
\begin{itemize}
  \item If the 0 is found in the last level of the conceptual tree,
  the 0 is simply replaced by a 1 to mark the new value of the cell
  and the update is complete.
  \item If the 0 is found in the upper levels of the conceptual
  tree, a new path must be created in the conceptual tree until
  the last level is reached. First, the 0 is replaced with a 1
  as in the previous case. Then, groups of \kk bits must be added
  in the lower levels. After replacing the 0 with a 1, a \irank operation
  is performed to compute the position where its children should be located.
  Then \kk 0 bits are added as children, and the one that ``covers''
  the position inserted is set to 1. The procedure continues
  recursively until it reaches the last level in the conceptual tree.
\end{itemize}

\begin{figure}[ht!]
\centering
   \includegraphics[width=0.9\textwidth]{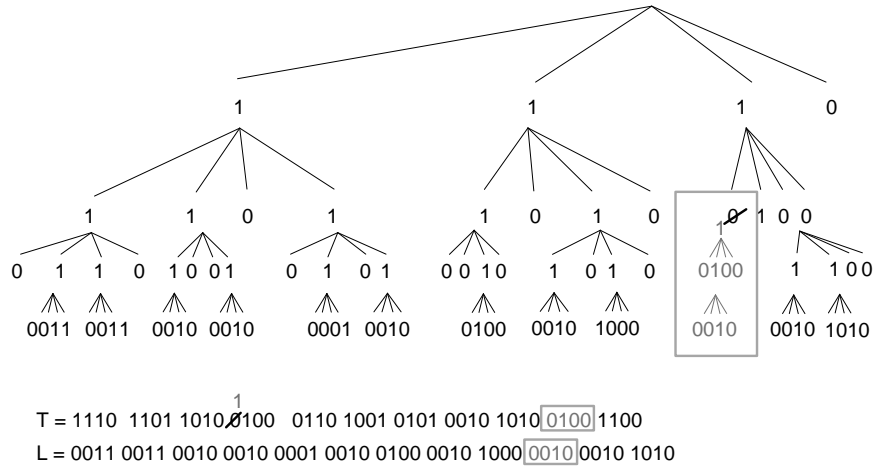}
   \caption{Insert operation in a \ktree: changes in the conceptual tree.}
   \label{fig:k2treeadd}
\end{figure}

Notice that there is still a third scenario corresponding to the
case where a 1 already exists in the cell to be inserted. However we
do not consider it, as in this case the element is already inserted
(and the appropriate path is already present as well), hence no
change is actually made in the representation.
Figure~\ref{fig:k2treeadd} shows an example of insertion in a
conceptual tree. At a given level in the conceptual tree a 0 is
found and replaced with a 1, and a new path is created adding \kk
bits to all the following levels. The new branch is highlighted in
gray and the changes in the bitmaps of the \ktree are also
highlighted.

To change a 1 into a 0 in the binary matrix, we need to set to 0 the
bit of the last level that corresponds to the cell. Then, the
current branch of the conceptual tree must be checked and updated if
necessary. First, we locate the position of the cell to be deleted
in the tree. The bit (node) corresponding to that cell is set to 0.
After this, we check the $\kk-1$ bits corresponding to the siblings
of that node. If at least one of the bits is set to 1 the procedure
ends. However, if all of them are set to 0, this means that there
are no 1s remaining in the current branch: we need to delete the
complete group of \kk 0 bits, move one level up in the conceptual
tree and set the bit corresponding to their parent node to 0. We
recursively repeat the same procedure upwards until a group of
non-zero \kk bits is found.

Summarizing the previous explanation, in order to support insertions
and deletions in the conceptual \dktree we only need to provide
three basic update operations in the dynamic bitmaps \ttree and
\ltree: flipping the value of a single bit, adding \kk bits at a
given position and removing \kk bits starting at a given position.
For example, Algorithm~\ref{alg:dk2tree-insert} shows the complete
process of insertion of new 1s in the matrix.

\newcommand{\leveldata}{\mathit{leveldata}}
\begin{algorithm}[ht!]
\caption{Insert operation} \label{alg:dk2tree-insert}
\begin{algorithmic}[5]
\scriptsize \Function{insert}{tree, r, c}
    \State $p \gets 0$
    \State  mode $\gets$ Search
    \For {$l \gets 0$ to tree.nlevels - 1}
        \State $p \gets \Call{computeChild}{p, r, c, l}$
        \State ($N_\ell$, $\bbefore$, $\obefore$) $\gets$ \Call{\emph{findLeaf}}{\ttree, $p$}
        \State $T_\ell \gets N_\ell.data$
        \If { mode = Search}
            \If {$\access(T_\ell, p-\bbefore) = 0$}
                \State $\Call{flip}{T_\ell, p - \bbefore}$
                \State mode $\gets$ Append
            \EndIf
        \Else
            \State $\Call{append4}{T_\ell, p-\bbefore}$
            \State $\Call{flip}{T_\ell, p-\bbefore}$
        \EndIf
        \State $rank \gets \obefore + \rank_1(T_\ell, p - \bbefore)$ \Comment $\rank_1(tree, p)$
        \State $p \gets rank \times \kk$
    \EndFor
    \State $l \gets tree.nlevels$
    \State $ p \gets p - tree.\ttree.length$
    \State $p \gets \Call{computeChild}{p, r, c, l}$
    \State ($N_\ell$, $\bbefore$) $\gets$ \Call{\emph{findLeaf}}{\ltree, $p$}
    \State $T_\ell \gets N_\ell.data$
    \If { mode = Search}
        \If {$\access(T_\ell, p-\bbefore) = 0$}
            \State $\Call{flip}{T_\ell, p - \bbefore}$
            \State mode $\gets$ Append
        \EndIf
    \Else
        \State $\Call{append4}{T_\ell, p-\bbefore}$
        \State $\Call{flip}{T_\ell, p-\bbefore}$
    \EndIf
\EndFunction

\end{algorithmic}
\end{algorithm}

To flip a single bit in \ttree or \ltree, we first retrieve the leaf
node $N_\ell$. The bit is changed in the bitmap of $N_\ell$ and its
local rank directory is updated (simply adding or subtracting 1 to
the value of the appropriate counter). Finally, if we are updating
\ttree, the $o$-counters in the entries followed in the path to
$N_\ell$ must be updated to reflect the change.

To add \kk bits at a given position in $N_\ell$, the \kk bits are
inserted in the bitmap of $N_\ell$ directly, and the counters in the
rank directory of $N_\ell$ must be updated accordingly (in this
case, all the counters from the position in $N_\ell$ where the
insertion has been done until the end of $N_\ell$ bitmap must be
updated, since the bitmap is displaced). After updating $N_\ell$,
the $b\dash$ and $o\dash$counters of its ancestors are also updated
accordingly (the $b\dash$ and $o\dash$counters are increased by \kk
and 1 respectively). Notice that we only update the $b\dash$ and
$o\dash$counters of the entries in the path to $N_\ell$ because only
those entries are affected by changes in the bitmap of $N_\ell$.

When a leaf of \ttree or \ltree reaches its maximum node size we
split it in two nodes, always keeping groups of \kk sibling bits in
the same leaf. This change is propagated to the parent of the leaf
node, causing the insertion of a new entry pointing to the new node
and updating the $b\dash$ and $o\dash$counters accordingly.
Eventually, internal nodes may also be split, evenly splitting their
entries in two new nodes.

To achieve better space utilization the \dktree can store nodes of
different maximum sizes. Given a base node size $B$, we allow a
number $e$ of partial expansions of the node before splitting it.
Hence, \ttree and \ltree may contain nodes of size $B$,
$B+\frac{B}{e+1}$, $\cdots$, $B+\frac{(e) B}{e+1}$ (class-0,
$\cdots$, class-$e$ nodes). If a node overflows, its contents are
reallocated in a node of the next class. If a fully-expanded node
overflows, it is split into two class-0 nodes.

\begin{figure}
\centering
   \includegraphics[width=0.9\textwidth]{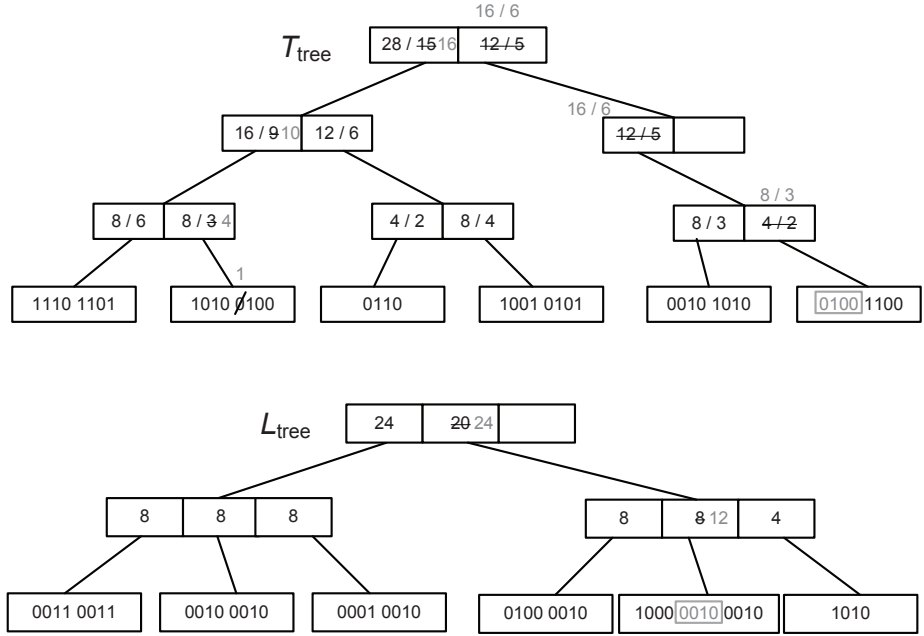}
   \caption{Insert operation in a \dktree: changes in the data structures.}
   \label{fig:k2treeadd2}
\end{figure}

\subsubsection{Changes in the rows/columns of the adjacency matrix}

The \dktree also supports the insertion of new rows/columns in the
adjacency matrix it represents, as well as deletion of existing
rows/columns.

The insertion of new rows/columns to the adjacency matrix is trivial
in many cases in the \dktree. Note that if the size of the matrix is
not a power of \k it is virtually expanded to the next power of \k
to represent it with a \ktree. Therefore, in a \ktree we usually
have unused rows and columns that can be made available. If the size
of the matrix is exactly a power of \k we can easily add new unused
rows expanding the matrix. To do this, we add a new root node to the
conceptual \ktree: its first child will be the current root and its
\kk-1 remaining children will be 0. This virtually increases the
size of the matrix from $\k^n \times \k^n$ to $\k^{n+1} \times
\k^{n+1}$. This operation simply requires the insertion of \kk bits
\textsf{1000} at the beginning of \ttree.

To delete an existing row/column, the procedure is symmetric to the
insertion. The last row/column of the matrix can be removed by
updating the actual size of the matrix and zeroing out all its
cells. Rows/columns at other positions may be deleted logically, by
adding these positions to a list of deleted rows and columns after
zeroing all their cells. This deleted rows/columns could be reused
later when new rows/columns are inserted, by just taking one from
the list.


\subsection{Analysis}

As previously stated in Section \ref{section:Introduction}, dynamic
representations of compact data structures are usually affected by a slowdown
factor that limits their overall efficiency when compared to a static
representation.
The static \ktree is essentially a LOUDS-based cardinal tree, and a dynamic
representation of a LOUDS tree has a slowdown factor of $\mathcal{O}(\omega)$ to
perform range queries~\cite[Section 12.5.2]{Nav16}. In the RAM model, we can
assume this to mean a slowdown of $\mathcal{O}(\log n)$ for any
dynamic representation of a LOUDS tree.

Note that in Section~\ref{improving} we described an optimization that takes
into account the access pattern in the \ktree (accesses to the tree are not random when
performing queries; they follow a well-defined pattern with many consecutive
accesses to close positions). This reduces the overall cost from the original
$O(C \log n)$ (where $C$ is the cost of the static representation, described
in Section~\ref{sec:ktree}) to $\mathcal{O}(C \log{\frac{n}{C}})$, improving
the result especially for costly queries.

Update operations over LOUDS cardinal trees require $\mathcal{O}(\log_k n)$
updates, in blocks of $k$ bits~\cite[Section 12.4.1]{Nav16}. The time required
for an update operation becomes $\mathcal{O}(\omega)$ if $k =
\mathcal{O}(\omega^2)$.
As this is a rather permissive value for $k$ in practice, we may consider the
update cost over the dynamic representation to be $\mathcal{O}(\log n \log_k n)$.

\section{Improved compression with a matrix vocabulary}
\label{sec:kdyn:structure:vocabulary}

\newcommand\vhash{\ensuremath{H}\xspace}
\newcommand\vempty{\ensuremath{V^{\mathrm{empty}}}\xspace}
\newcommand\vfreq{\ensuremath{F}\xspace}

Recall from Section~\ref{sec:k2tree:improvements} that the \ktree
space results can be improved using a matrix vocabulary to compress
the bitmap $L$. This improvement replaces the plain bitmap $L$ with
a sequence of variable-length codes and a matrix vocabulary. In this
section we introduce a similar proposal for \dktrees. In this
proposal, the leaves in \ltree will store a sequence of
variable-length matrix identifiers encoded using ETDC~\cite{etdc,
etdcj}.

Note that the management of a matrix vocabulary is much more complex
in the \dktree: we should be able to add and remove entries from the
vocabulary, as well as efficiently check whether a given matrix
already exists in the vocabulary. Also, when \ltree stores a
sequence of codewords, the actual number of bits and ones in a leaf
is no longer the same as the \emph{logical} values stored in
$b\dash$ and $o\dash$counters of its parent entry. This does not
affect the tree structure because the actual size of each leaf node
is stored in the \emph{size} field of its header, and this is the
value used to determine when to expand or split a leaf, while the
values in the counters are still used as before to access the
appropriate leaf.

To store the matrix vocabulary we built a simple
implementation that stores a hash table \vhash to look up matrices.
An array $V$ stores the position in \vhash where each matrix is
stored. Finally, we add another array \vfreq that stores the
frequency of each matrix. An additional list \vempty stores the
codewords that are not being currently used.

\begin{figure}
\centering
   \includegraphics[width=0.80\textwidth]{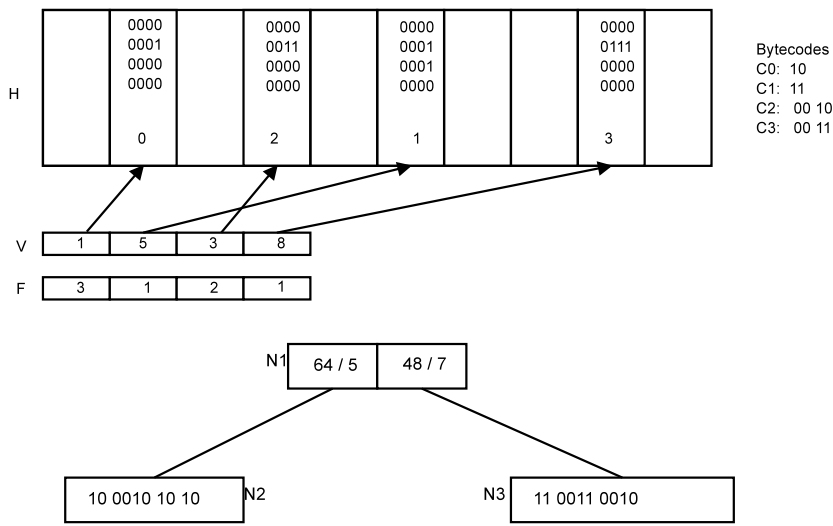}
   \caption{Dynamic vocabulary management: vocabulary (top) and \ltree (bottom).}
   \label{fig:vocabulary}
\end{figure}

Figure~\ref{fig:vocabulary} shows an example with a complete
vocabulary representation. The leaves of \ltree (bottom, nodes N2
and N3) store a sequence of variable-length codes represented with
ETDC (we consider 2-bit chunks in this simplified example). Notice
that the $b\dash$ and $o\dash$counters in internal node N1 still
refer to the logical size of the leaf: the entry pointing to N2
marks it as containing 64 bits (4 submatrices of size $4\times 4$)
and 5 ones. The submatrices are stored in a hash table that contains
for each matrix its offset in the vocabulary (that can be easily
translated into its ETDC codeword). $V$ points to the entry in $H$
for each vocabulary codeword, and $F$ stores its frequency.

To \iaccess a position in \ltree when using a matrix, we obtain a \emph{logical} offset in the leaf from \findleaf.
To retrieve the actual bit, we sequentially traverse the sequence of
variable-length codes in the leaf. When we find the code that
contains the desired position, we translate the codeword into an array index, and $V$
is used to retrieve the actual matrix in $H$. For example, suppose we want
to access position 21 in the example of Figure~\ref{fig:vocabulary}.
Our \findleaf operation would take us to node N2, offset 21. To
obtain the actual matrix we would traverse the codes in N2, taking
into account the actual size of each submatrix (16 bits), so our
offset would be at position 5 in the second submatrix. We go over
the code \texttt{10} and find the second code \texttt{0010}. To find
the actual matrix, we convert this code to an offset (2) and access
$V[2]$ to locate the position in $H$ where the matrix is actually
stored (3, second non-empty position). Finally, in $H$ we can access
bit 5 in the matrix bitmap (0).

The main difference with a static implementation is the need to
sequentially traverse the list of variable-length codes. We can
reduce the overhead of this sequential traversal adding to the
leaves of \ltree a set of samples that store the actual offset in
bytes of each $S_L$-th codeword. The idea is similar to the sampling
used for \irank in the leaves of \ttree. With this improvement, to locate a given position we can simply use
the samples to locate the previous sampled codeword and then start
the search from the sampled position.

\subsection{Update operations}

Update operations in \ltree, when using a matrix vocabulary, require us to add, remove or modify
variable-length codes from the leaves of \ltree. All the update
operations start by finding the real location in the node where the
variable-length code should be added/removed/modified.

To insert groups of \kk bits we need to add a new codeword. The
matrix corresponding to the new codeword is looked up in \vhash,
adding it to the vocabulary if it did not exist and increasing its
frequency in \vfreq. Then, the codeword for the matrix is inserted
in the leaf, updating the counters in the node and its ancestors.

To remove groups of \kk bits in a leaf node of \ltree a codeword
must be removed: we locate the codeword in \ltree, decrease its
frequency in \vfreq and then we remove the code from the leaf node,
updating ancestors accordingly. If the frequency of the codeword
reaches 0, the corresponding index in $V$ is added to \vempty. When
new entries must be added to the vocabulary \vempty will be checked
to reuse previous entries and new codes will only be used when
\vempty is empty.

To change the value of a bit in \ltree we need to replace existing
codewords. First, the matrix for the current codeword is retrieved
in \vhash and its frequency is decreased in \vfreq. We look up the
new matrix in \vhash. Again, if it already existed, its frequency is
increased, and if it is new, it is added to \vhash and its frequency
set to 1. Then, the codeword corresponding to the new matrix is
inserted in \ltree replacing the old codeword.

Following the example of Figure \ref{fig:vocabulary}, suppose that
we need to set to 1 the bit at position 21 in \ltree. \findleaf
would take us to N2, where we have to access the second bytecode
\texttt{00 10} at offset $5$. This bytecode (C2) corresponds
to offset 2 in the ETDC order. We would access $V[2]$ to retrieve
the corresponding matrix.
The operation would require us to
transform the matrix as follows:
\[
    \begin{array}{c}
        0 0 0 0 \\
        0 0 1 1 \\
        0 0 0 0 \\
        0 0 0 0 \\
    \end{array}
    \longrightarrow
    \begin{array}{c}
        0 0 0 0 \\
        0 1 1 1 \\
        0 0 0 0 \\
        0 0 0 0 \\
    \end{array}
\]

The new submatrix already exists, at position 3 in $V$ with frequency 1. Hence, we would need to update
the leaf node replacing the old codeword \texttt{00 10} with the new
codeword C3: \texttt{00 11}. The vocabulary would also be updated, decreasing the frequency of C2 to 1 and increasing the frequency of C3 to 2.

\subsection{Handling changes in the frequency distribution}

The compression achieved by the matrix vocabulary depends heavily on
the evolution of the matrix frequencies. As the distribution of the
submatrices changes the efficiency of the variable-length codes will
degrade. The simplest approach to mitigate this problem is to use a
precomputed vocabulary from a fraction of the matrix to obtain a
reasonably good frequency distribution.

To obtain the best compression results, when the frequency of a
submatrix changes too much its codeword should also be changed to
obtain the best compression results. This is a process similar to
the vocabulary adjustment in dynamic ETDC (DETDC \cite{detdc,
dletdc}). However, in a \dktree, to change the codeword of a
submatrix, we must also update all the occurrences of the codeword
in the leaves of \ltree. Therefore, a space/time tradeoff exists
between vocabulary size and update cost.

To maintain a compression ratio similar to that of static \ktrees we
can use simple heuristics to completely rebuild \ltree and the
matrix vocabulary: rebuild  every $p$ updates or count the
number of inversions in \vfreq. To rebuild \ltree, we must sort the
matrices in $H$ according to their actual frequency and compute the
new optimal codes for each matrix. Then we have to traverse all the
leaves in \ltree from left to right, replacing the old codes with
optimal ones. Notice that the replacement can not be executed
completely in place, because the globally optimal codes may be worse
locally, but the complete process should require only a small amount
of additional space. After \ltree is rebuilt, the old vocabulary is simply
replaced with the optimal values.

\newcommand\vperm{\ensuremath{{\mathit{VP}}}\xspace}
\newcommand\voctop{\ensuremath{\mathit{Top}}\xspace}

Instead of using the simple heuristics to rebuild the matrix
vocabulary, we can keep track of how good the current compression
is. To guarantee that the compression of \ltree is never too far
from the optimum, we can keep track of the actual optimum
vocabulary. To do this, we propose an enhanced vocabulary
representation, similar to the adaptive encoding used in DETDC. In
our case it would be unfeasible to change the actual vocabulary each
time the length of a codeword changes, but we store the optimal
vocabulary to know exactly the amount of space that would be gained
using an optimal vocabulary.

\begin{figure}[ht!]
\centering
   \includegraphics[width=0.9\textwidth]{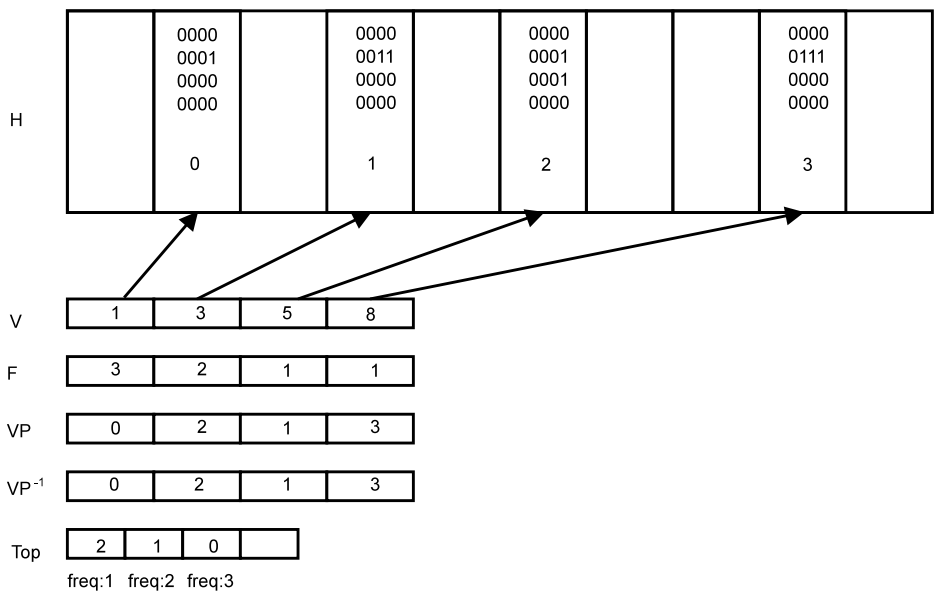}
   \caption{Extended vocabulary to keep track of the optimal codewords.}
   \label{fig:vocabulary-complete}
\end{figure}

To this aim we store, in addition to \vhash and \vfreq, a
permutation \vperm between the current vocabulary and the optimal
one: $\vperm(i)$ gives the optimal position of the codeword at
offset $i$, while $\vperm^{-1}(i)$ gives the current position given
the optimal position. This permutation will allow us to keep $V$ and
\vfreq always sorted in descending order of frequency (that is,
according to the optimal vocabulary).
Figure~\ref{fig:vocabulary-complete} shows the data structures
required to represent the same vocabulary of Figure
\ref{fig:vocabulary} using the new method.

In this representation, we obtain the matrix for a codeword as in
the previous version; the only difference is that we first obtain
the offset of the codeword in the optimal vocabulary and then we use
$\vperm^{-1}$ to compute the offset in the current vocabulary. Also,
to find the matrix for a given codeword we compute the optimal
offset for the codeword using \vperm and then use $V$ to find the
position of the matrix.

To keep track of the changes in frequencies, we also build an array,
\voctop, that stores, for each different frequency, the position in
$V$ of the first codeword with that frequency. The array \voctop is
used to swap codewords when frequencies change, as it is performed
in DETDC. If the frequency of a matrix at index $i$ in $V$ changes
from $f$ to $f+1$, the new optimum position for it would be the
position $\voctop[f]$. The indexes in $V$, \vperm and $H$ would be
updated to reflect the change. The case when the frequency of a
matrix decreases from $f$ to $f-1$ is symmetric: we swap the current
position of the matrix with $\voctop[f-1]+1$, updating the
corresponding indexes in $F$ and \vperm.

\newcommand\vthres{$\mathit{reb_{\ltree}}$}

The use of the extra data structures allows us to control precisely
how much space is being wasted at any moment. This means that we can set a threshold
$\mathit{reb_{\ltree }}$ and rebuild \ltree when the ratio
$\frac{\mathit{size_{cur}}}{\mathit{size_{opt}}}$ surpasses it.

In order to physically store all the data structures required for
the dynamic vocabulary, we resort to simple data structures that can
be easily updated. \vhash is a simple hash table backed by an array.
$V$ and \vfreq are extendible arrays. If we want to set the
threshold \vthres we need additional data structures for \vperm and
\voctop. \voctop can be implemented using an extendible array and
two extendible arrays can store \vperm and its inverse. The goal of
these representations is to provide efficient update times (recall
that each update operation in the \dktree will always lead to a
change in \ltree, that will cause at least one frequency change in
the vocabulary).

\section{Experimental Evaluation} \label{sec:kdyn:experiments}

In this section we experimentally test the efficiency of the \dktree in order to
demonstrate its capabilities to answer simple queries in space and
time close to those of the static \ktree data structure.

First, we will study the different parameters of the \dktree and
their effect in compression and query efficiency. Then, we will show
the efficiency when compared to the equivalent static data structure
in the original application domain of \ktrees: Web graph
representation. Finally, Section \ref{sec:kdyn:experiments:rdf} will
be devoted to describe the application of \dktrees to the
representation of RDF datasets, where the \ktree has been already
used and dynamic operations are of interest. In this context, we
will compare our representation with state-of-the-art alternatives
including a similar static approach based on \ktrees.

All the experiments in this article were run on an AMD-Phenom-II X4 955@3.2 GHz,
with 8GB DDR2 RAM. The operating system is Ubuntu 12.04. All our code is
written in C and compiled with gcc version 4.6.2 with full
optimizations enabled.

\begin{figure}[th!]
    \centering
    \begin{tabular}{c}
    \vspace{-5mm}
        \subfigure{
        \includegraphics[angle=270,width=0.5\textwidth]{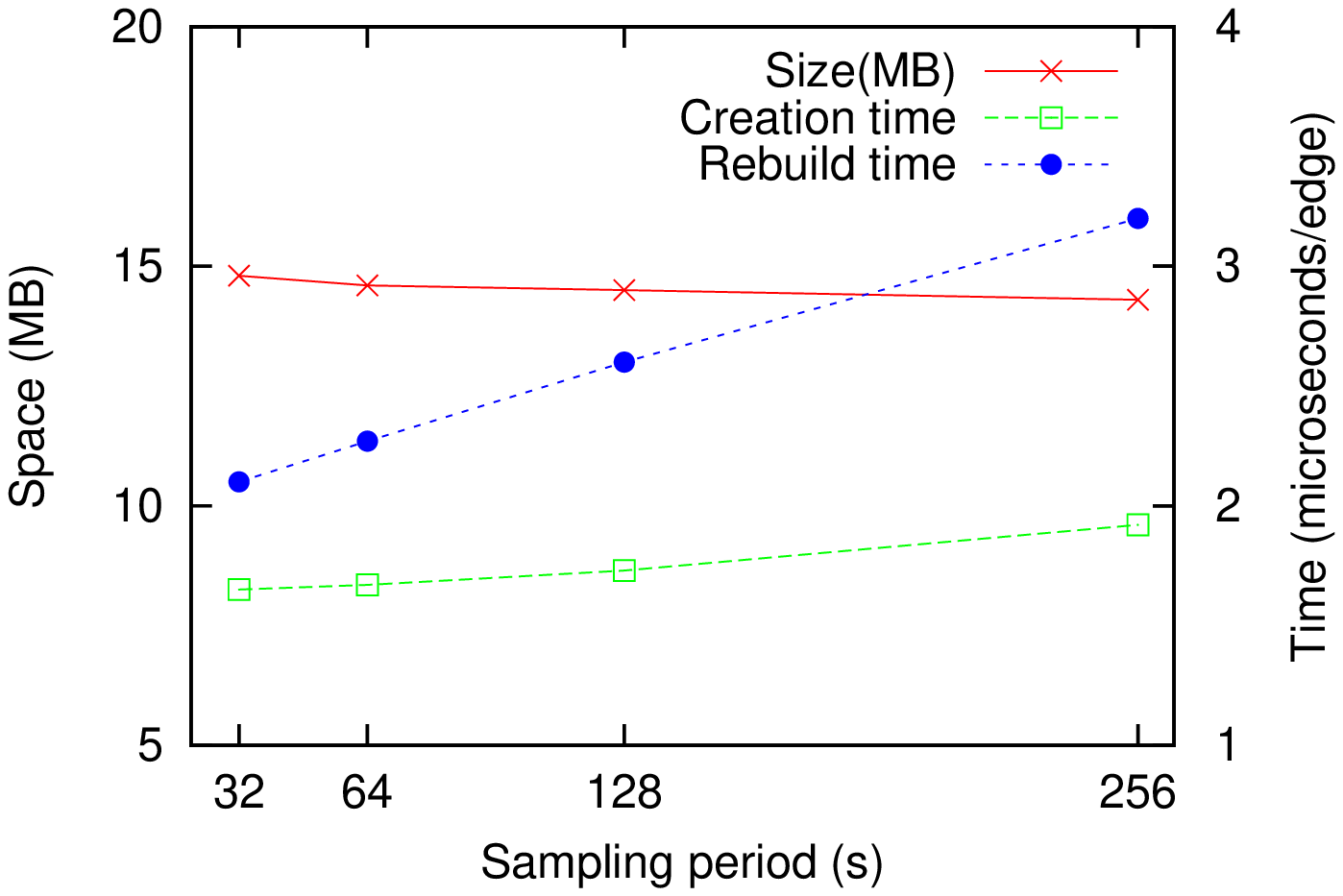}
        }
    \vspace{-5mm}
        \subfigure{
        \includegraphics[angle=270,width=0.5\textwidth]{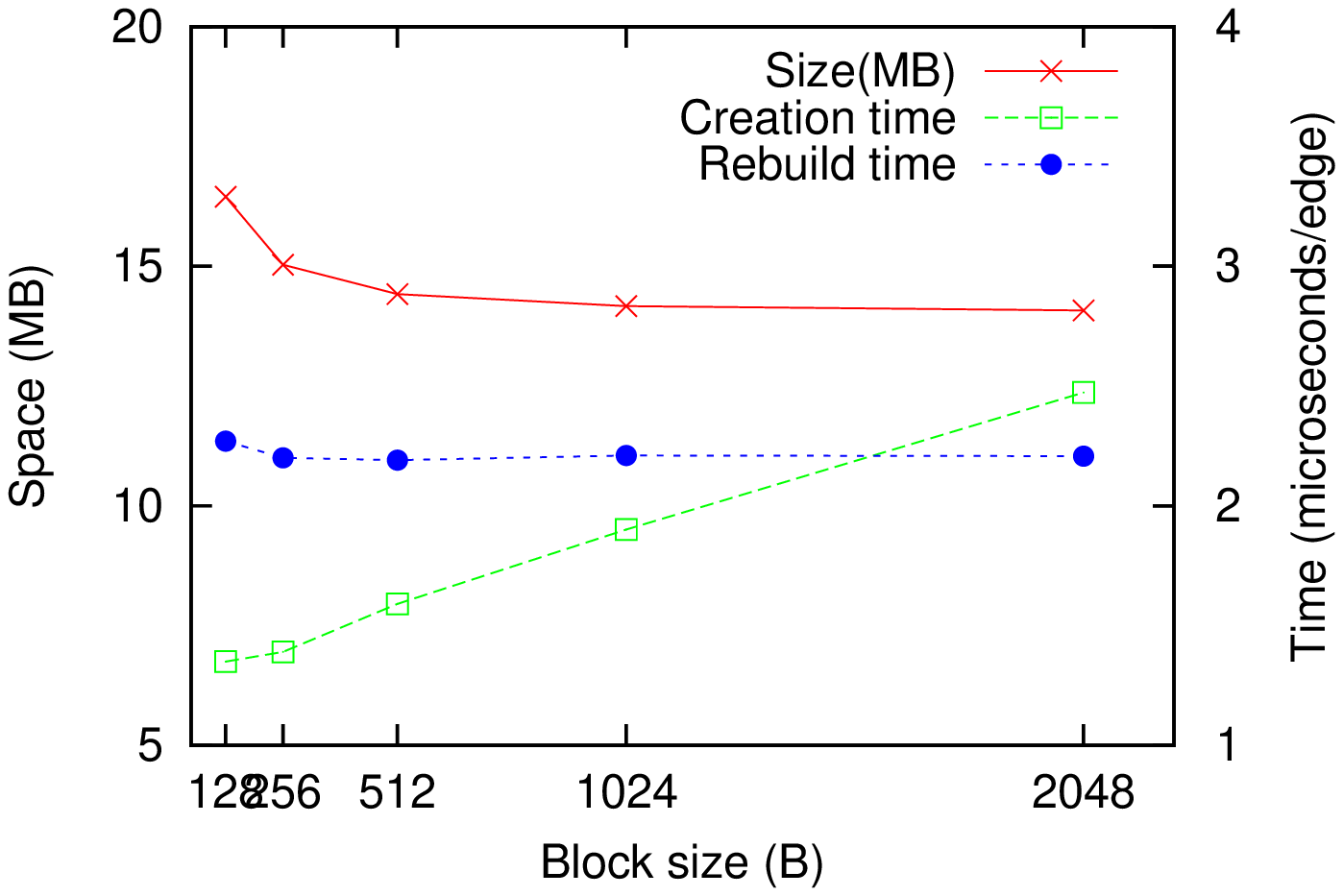}
        }\\
    \vspace{-5mm}
        \subfigure{
        \includegraphics[angle=270,width=0.5\textwidth]{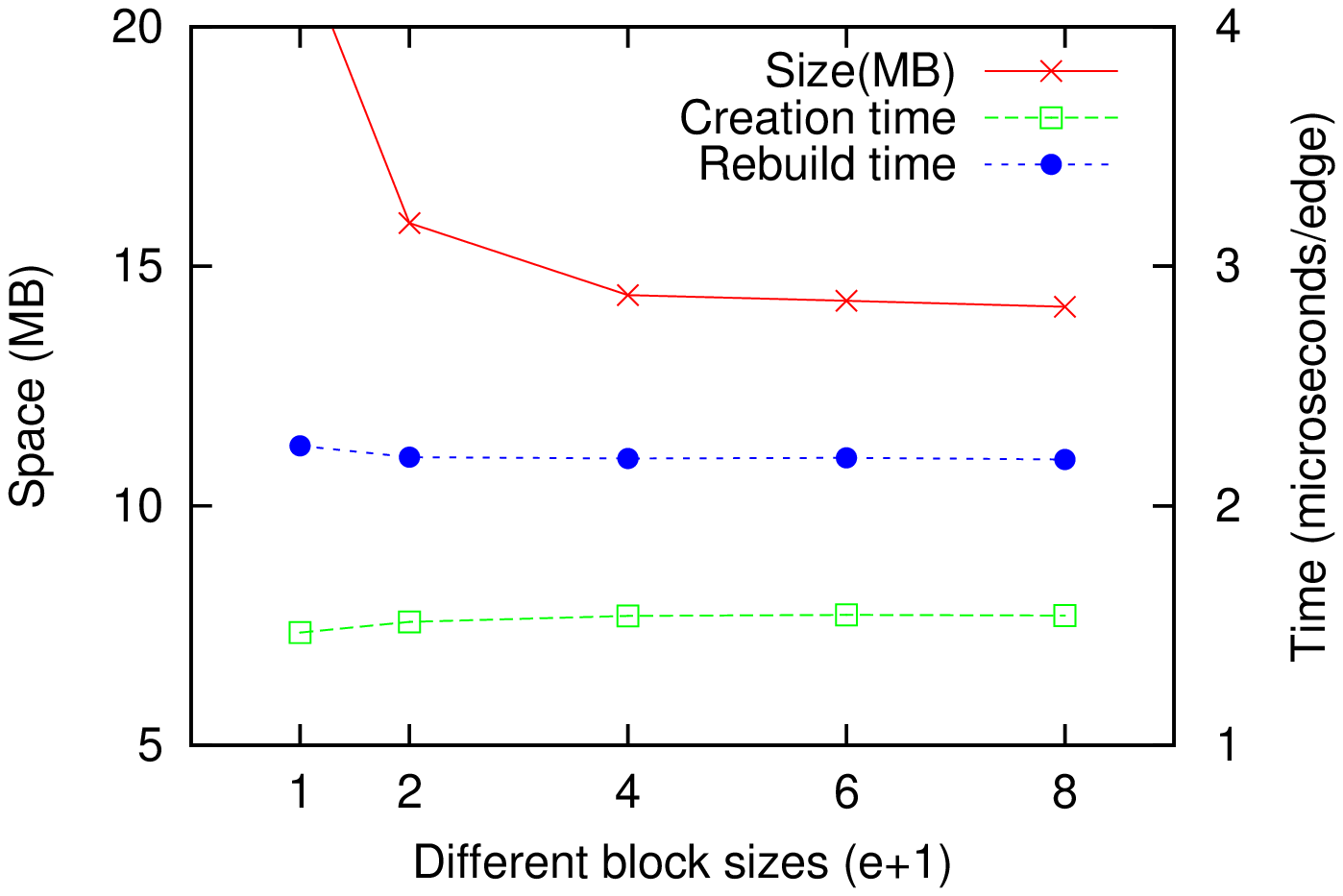}
        }
    \end{tabular}
    \vspace{5mm}
   \caption{Evolution of space/time results of the \dktree changing the parameters $s$, $B$ and $e$.}
   \label{fig:evolution}
\end{figure}

\subsection{Parameter tuning}\label{sec:kdyn:experiments:tuning}

The main parameters used to settle the efficiency of the \dktree are
the sampling period $s$ in the leaves of \ttree ($S_T$) and \ltree
($S_L$), the block size $B$ on the nodes and the number of partial
expansions $e$; the value $k'$ in the last level is also important
when a matrix vocabulary is used for $L$. We will first focus on the
effect of the first parameters, and then study the effect of the
matrix vocabulary independently.

We use for our experiments a Web graph dataset,
\emph{eu-2005}\footnote{Dataset from the \emph{WebGraph} project,
that comprises some Web graphs gathered by UbiCrawler \cite{BCSV04}.
These datasets are made available to the public by the members of
the \emph{Laboratory for Web Algorithmics}
(\texttt{http://law.di.unimi.it}) at the \emph{Universit\`a Degli
Studi Di Milano}.}, a small graph with 19 million edges. The results
of parameter tuning are similar for other datasets used in following
sections. We do not use a matrix vocabulary in this first example.
Figure~\ref{fig:evolution} shows the evolution of the \dktree size
and the creation and rebuild time depending on each parameter. The
\dktree size, in MB, is the overall memory used by the data
structure. The creation time is the time to build the \dktree from a
plain representation inserting each edge separately, so it provides
an estimation of the average update time of the structure. Finally,
the rebuild time is the time to retrieve all the 1s in the adjacency
matrix in a single range query covering the complete matrix, and it
is shown as a rough estimation of the expected evolution of query
times. Both times are shown in microseconds per element
inserted/retrieved.




The top-left plot in Figure~\ref{fig:evolution} shows the results
obtained for different values of the sampling interval $s$, with
fixed $B=512$ bytes and $\#classes=e+1=4$, but the tradeoff is
similar for different values. Smaller values of $s$ increase the
size of the trees slightly, but a considerable reduction in query
time is obtained. Additionally, update operations can also be
improved by using smaller values of $s$. Even though blocks with
more samples are more costly to update when their contents change,
the recomputation of the samples is only performed if the node
contents actually change. On the other hand, the \emph{rankLeaf}
operation must always be performed at all the levels of the
conceptual tree, and its cost is significantly reduced when using
smaller values of $s$. Therefore, a small sampling period can be
used to obtain faster access times with only a minor increase in the
size of the \dktree.

The top-right plot in Figure~\ref{fig:evolution} shows the results
for different values of $B$, for fixed $s=128$ bytes and
$\#classes=e+1=4$. The block size $B$ provides a clear space/time
tradeoff: small values of $B$ yield bigger \dktrees due to the
amount of overhead to store many smaller nodes, while larger values
of $B$ make updates become more costly. Query times are not very
different depending on $B$ for usual values. In our experiments we
will choose values of $B=256$ or $B=512$ to obtain good space
results with small penalties in update times.

The bottom plot of Figure~\ref{fig:evolution} shows the evolution with $e$ for fixed $s=128$ and $B=512$. If we
use a single block size ($e+1 = 1$), the node utilization is low and
the figure shows poor space results, but even for a relatively small number
of block sizes the space results improve fast with only minor changes in the creation and rebuild times.

After measuring the basic parameters of the \dktree data structure,
we focus on the analysis of the matrix vocabulary variant. We build
a \dktree for several Web graph datasets with and without a matrix
vocabulary and for different values of the parameter $\k'$. We
compare the static and dynamic representations in two different Web
graph datasets\footnote{Again, datasets obtained from the
\emph{WebGraph} project (\texttt{http://law.di.unimi.it}).}: the
\emph{indochina-2004} dataset, with 200 million edges, and the
\emph{uk-2002} dataset, with 300 million edges. In all cases we
built a hybrid variant of the \ktree or the \dktree, with $\k=4$ in
the first 5 levels of decomposition and $\k=2$ in the remaining
levels. For the variants with matrix vocabulary, we test the values
$\k'=4$ and $\k'=8$. In the \dktree we choose a block size $B=512$,
$e=3$ (4 different block sizes) and $s=128$. The static \ktree
representation uses a sampling factor of 20 for its rank data
structures, hence requiring an additional 5\% space.

We use in the \dktree the most complex version of the matrix vocabulary, that keeps track of the optimum
vocabulary and rebuilds the complete vocabulary when the total size
is 20\% worse than the optimum. Additionally, we set a
threshold of 100 KB for the size of \ltree, so that the vocabulary
is only checked (and rebuilt if necessary) when \ltree reaches that
size. We also consider in the \dktree two different
scenarios: the space required by the simplest version of the
vocabulary (\emph{dynamic}) and the total space required to keep track of the optimum vocabulary
(\emph{dynamic-complete}).

\begin{figure}[H]
    \centering
    \includegraphics[width=\textwidth]{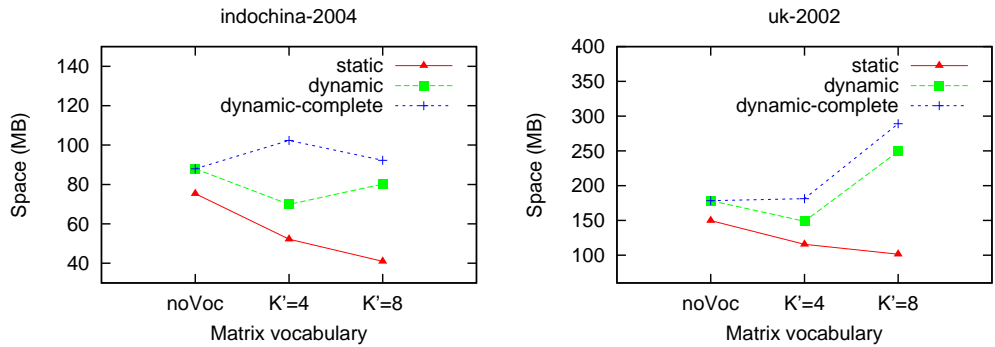}
    \caption{Space utilization of static and dynamic \ktrees with different matrix vocabularies in Web graphs.}
    \label{fig:dktree:vocabulary}
\end{figure}

Figure~\ref{fig:dktree:vocabulary} shows the evolution of the space
utilization for both datasets required by original \ktrees (static)
and a \dktree (dynamic). Note that the \dktree space utilization is always close to
that of the static \ktree when no matrix vocabulary is used
($noVoc$). The space overhead of \dktrees, around 20\%, is mostly
due to the space utilization of the nodes of \ttree and \ltree.

Static \ktrees obtain better compression for larger values of \k',
reaching their best space utilization when $\k'=8$. On the other
hand, the \dktree improves its space results only for small \k'.
Note also that the variant that keeps track of the optimum
vocabulary ($dynamic-complete$) is always bigger than the simpler
approach. In fact, in our experiments the graphs were only rebuilt
once, when the size of \ltree reached the threshold, showing that
once a small fragment of the adjacency matrix has been built the
resulting matrix vocabulary becomes good enough to compress the
overall matrix with a relatively small penalty in space. Hence, the
size of the additional data structures required to keep track of the
optimum vocabulary is higher than the space reduction of the vocabulary itself. 
Considering these results, a simpler strategy to maintain a ``good''
matrix vocabulary (such as using a predefined matrix vocabulary
extracted from experience or simply rebuilding after $x$ operations)
may be the best approach in many domains. On the other hand, the
strategy to keep track of the optimum vocabulary could still be of
application in domains where the relative size of the matrix
vocabulary is expected to be of small size.

\subsection{Query and update times}
\label{sec:kdyn:experiments:wg}

In this section we extend the previous analysis of the \dktree measuring the efficiency of our
proposal in terms of query and update times.

To measure the query efficiency, we focus on the representation of
Web graphs, the original application domain of the static \ktree. We
choose the most usual query in this domain, namely, the
\emph{successor} query that asks for the direct neighbors of a
specific node (all the cells with value 1 in a specific row of the
adjacency matrix). For each dataset we run successor queries for all
the nodes in the dataset and measure the average query times in
$\mu$s/query. 

\begin{figure}[H]
    \centering
    \includegraphics[width=\textwidth]{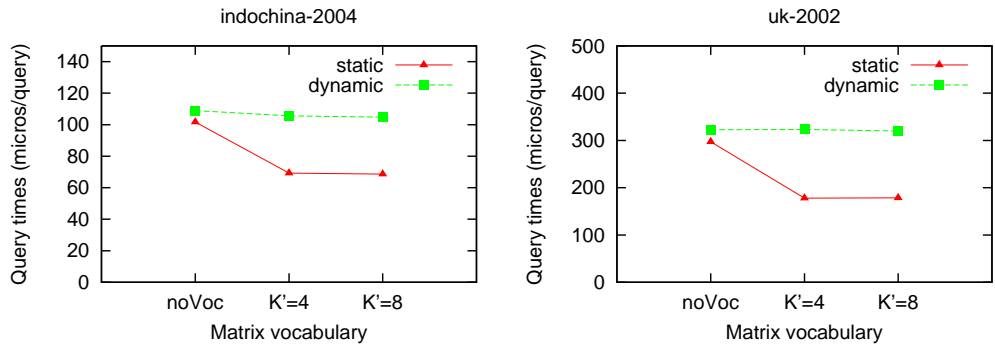}
    \caption{Time to retrieve the successors of a node in a static \ktree and a \dktree. Query times in $\mu$s/query.}
    \label{fig:dktree:neighbors}
\end{figure}

As shown in Figure~\ref{fig:dktree:neighbors}, the \dktree is always
slower than a static representation. Comparing the \dktree version
that obtained the best space results ($\k'=4$) with the best static
\ktree version ($\k'=8$), the \dktree is 50-80\% slower than the
static data structure. This difference in query times is
significant, but for specific scenarios where update operations are
frequent, the \dktree turns out to be a reasonable solution especially if we
consider the slowdown that affects any dynamic representation, and the
limitations of its static version.

The cost of update operations in the \dktree depends on several
factors, such as the choice of parameters $B$ and $s$. The
characteristics of the dataset also have a great influence in its
\ktree and \dktree representation, since the clusterization of 1s
lead to a better compression of the data. In the \dktree, the
clusterization of 1s and the sparsity of the adjacency matrix also
affect update times: when new 1s must be inserted and they are far
apart from any other existing 1, the insertion operation must insert
\kk bits in many levels of the conceptual \ktree, which increases
the cost of the operation. Therefore, insertion costs are expected
to be higher on average when datasets are very sparse.

To measure this effect of the distance between 1s on update costs, 
we choose to use synthetic datasets. Since we aim to 
evaluate the insertion cost depending on the level of the conceptual 
tree where that insertion is performed, synthetic data allow us 
to specifically control that without depending on other features 
of specific real Web graphs that were used instead. We create a set of very 
sparse synthetic datasets. In them, 1s are inserted 
every $2^d$ rows and $2^d$ columns,
so that the \ktree representation has a unary path of length $d$ to
each edge. Table~\ref{tab:dktree:synthDatasets} shows a summary with
the basic information of the datasets. We choose the separation for
the different dataset sizes so that all the datasets have the same
number of edges (4,194,304).

\begin{table}[ht!]
\centering
\begin{tabular}{|c | r | r | r |}
\hline
Dataset     &   \#rows/columns  &   Separation between 1s ($2^d$)   &   \# \ktree levels    \\
\hline
synth\_22   &   4,194,304       &   2,048 ($d=11$)          &   22  \\
synth\_24   &   16,777,216      &   8,192 ($d=13$)          &   24  \\
synth\_26   &   67,108,864      &   32,768  ($d=15$)        &   26  \\
\hline
\end{tabular}
\caption{Synthetic sparse datasets used to measure insertion costs.}
\label{tab:dktree:synthDatasets}
\end{table}

We measure the insertion cost with these datasets, depending on the
number of levels $\ell$ that must be created in the conceptual
\ktree to insert the new 1. We compare the insertion costs for $\ell
\in [0,10]$. For each dataset and value of $\ell$, we create a set
of 200,000 cells of the matrix that require exactly $\ell$ new
levels in the conceptual tree. In this experiments we use a simple
setup with a single block size. Additionally, we compute the cost to
\emph{query} the new cells over the unmodified synthetic datasets.
These queries are an approximation of the insertion cost that is
actually due to locating the node of the conceptual \ktree where we
must start the insertion.

\begin{figure}[ht!]
    \centering
    \includegraphics[width=0.48\textwidth]{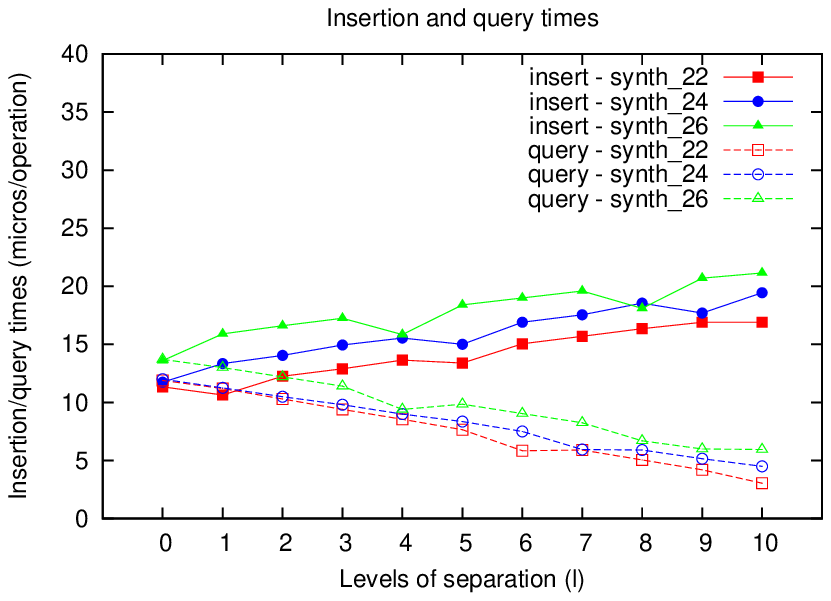}
    \includegraphics[width=0.48\textwidth]{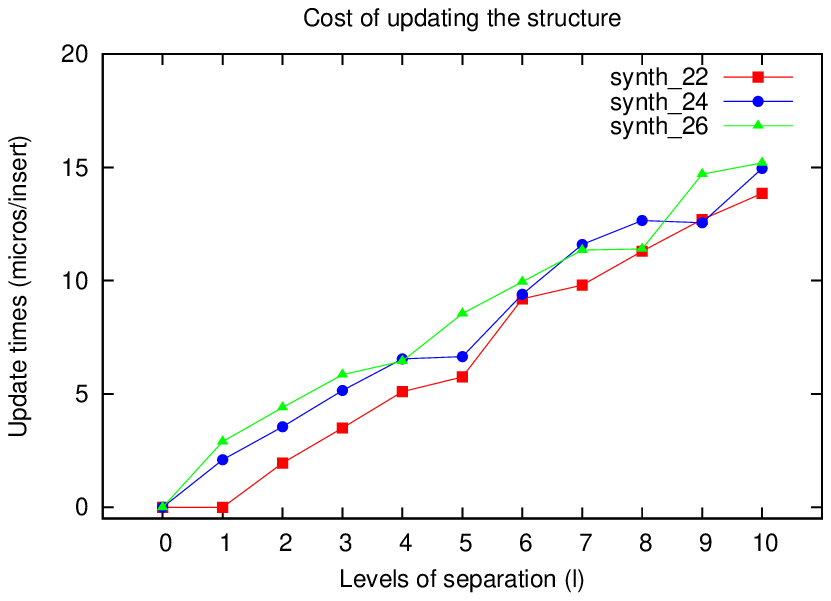}
    \caption{Insertion and query times (left) and estimation of update cost (right) in the \dktree varying $\ell$.}
    \label{fig:dktree:insert}
\end{figure}

Figure~\ref{fig:dktree:insert} (left) shows the evolution of
insertion and query times, in $\mu s/edge$, in the different
datasets. When $\ell$ is small (new 1s are inserted very close to
existing 1s), insertion and query times are almost identical. As
$\ell$ increases the insertion cost becomes higher while query times
become lower because the first 0 in the conceptual tree (the node
where insertion should start) is found in upper levels of the tree.
Figure~\ref{fig:dktree:insert} (right) shows an estimation of the
actual cost devoted to update the tree depending on $\ell$, and
computed by subtracting query times from insertion times. Notice
that it is 0 for small $\ell$ and steadily increases with $\ell$.

Our overall results suggest that insertion times in the \dktree can
be very close to query times if the represented dataset has some
properties that are also desirable for compression, particularly the
clustering of 1s in the binary matrix. The evolution of insertion
times shows that insertions at the upper levels of the tree may be
several times more costly than insertions in the lower levels of the
tree. However, insertions in the upper levels of the tree should be
very infrequent in most of the datasets where a \dktree will be
used, since compression in \ktrees also degrades when matrices have
no clusterization at all.

\section{Representation of RDF databases}
\label{sec:kdyn:experiments:rdf}

RDF (\emph{Resource Description Framework}) \cite{RDF:04} has become
an increasingly popular language recommended by the W3C for the
description of facts in the Web of Data. It follows a graph-based
data model, in which the information is represented as a set of
triples. Each triple is an edge of a labeled graph, and represents a
property of a resource using a subject \emph{S} (element of
interest, or source node), a predicate \emph{P} (property of the
element, or edge label), and an object \emph{O} (value of the
property, or target node). These (S,P,O) triples can be queried
using a standard graph-matching language named SPARQL
\cite{SPARQL:13}. This language is built on top of triple patterns,
that is, RDF triples in which each component may be a variable
(variables are preceded, in the pattern, by the symbol $?$):
$(S,P,O)$, $(?S,P,O)$, $(?S,?P,O)$, etc. Yet, more complex queries
can be created by joining sets of triple patterns.

Many different proposals have appeared in recent years to
efficiently store and query RDF datasets. Many of these so called
``RDF stores'' are based on relational databases, creating specific
database structures to store the triple patterns in the RDF
data~\cite{SWSTORE,SIDIVP}. Other particular solutions rely on
specific data structures designed to compactly store the data while
allowing efficient query operations~\cite{RDF3X, Fernandez201322,
CPE:CPE2840, UDGW16}.

RDF datasets can be built from snapshots of data and therefore
stored in static form. However, in many cases new information is
continuously appearing and must be incorporated into the dataset to
keep it updated. In these cases, a purely static representation of
the dataset is unfeasible, since a way to efficiently update the
contents in the RDF dataset is needed. Most of the relational
approaches for RDF storage can handle update operation. However,
specific compact representations usually lack the same flexibility,
but still some proposals exist, like X-RDF-3X~\cite{XRDF3X}, a
dynamic evolution of the existing multi-indexing native solution
RDF-3X.

A representation of RDF datasets based on \ktrees, called \ktriples,
has been presented in~\cite{Alvarez15}. This representation uses a
collection of \ktrees to represent the triples corresponding to each
predicate in the RDF dataset. This representation was proved to be
very competitive in space and query times with state-of-the-art
alternatives. However, this proposal was limited to a static context
due to the static nature of \ktrees.

In this section we propose a dynamic representation of RDF datasets based
on \dktrees, that simply replaces static \ktree representations with
a \dktree per predicate. We aim to demonstrate that a representation
based on the \dktree can obtain query times close to the static
\ktriples approach, but more importantly that the \dktree
representation provides the basis to perform update operations on
the RDF dataset, which are actually expected operations in real
applications.

\subsection{Our proposal}

Our proposal simply replaces the static \ktree representation in
\ktriples with a \dktree per predicate. We consider a partition of
the RDF dataset by predicate, and build a \dktree for each predicate
in the dataset. For each predicate we consider a matrix storing all
relations between subjects and objects with that predicate. All
matrices will contain the same rows/columns, where subject-objects
(elements that appear as both subjects and objects in any triple)
will be located together in the first rows/columns of the matrix and
the remaining rows (columns) of the matrices will contain the
remaining subjects (objects)\footnote{Notice that we follow the same
subject-object arrangement than that used by the static approach,
thus assuming the use of a similar strategy to create the vocabulary
of terms.}.


Our proposal based on \dktrees aims to solve the representation of
the structural part of an RDF dataset. We assume that additional
data structures must be used to store vocabularies of subjects,
objects and predicates and map them with rows/columns of the
matrices represented by \dktrees. The creation of a dynamic and
efficient dictionary representation to manage large collections of
URIs and literal values is a complex problem, since the vocabulary
may constitute a large part of the total size of an RDF dataset
\cite{dicrdf}.

As previously stated, the main query operations in RDF datasets are
based on triple-pattern matching. These triple patterns can be
easily translated into a collection of simple queries in one or more
of the \ktrees used to store the complete RDF dataset. Hence, our
proposal can directly answer all triple pattern queries: $(S,P,O)$
and $(S,?P,O)$ queries are actually cell retrieval queries
(involving one \dktree or all the \dktrees in the collection,
respectively), $(S,P,?O)$, $(S,?P,?O)$, $(?S,P,O)$ and $(?S,?P,O)$
are row/column queries and $(?S,P,?O)$ is a full range retrieval
query that asks for all the cells in a \dktree.

Triple patterns can usually be joined to build more complex queries.
Join operations involve matching multiple triple patterns with a
common element. For instance, the query
$(?S,P_1,O_1)\bowtie(?S,P_2,O_2)$ represents all the subjects that
relate to $O_1$ with predicate $P_1$ and to $O_2$ with predicate
$P_2$. The join variable is marked with an $?$, and is a common
element to both triple patterns. In static \ktriples, three
different strategies were proposed to solve triple patterns, and all
of them can be also used in our \dktrees:
\begin{itemize}
  \item \emph{Independent evaluation} separates any join operation in two
  triple pattern queries and a simple merge that intersects the
  results of both queries. The adaptation to \dktrees is trivial using the basic operations explained.
  \item \emph{Chain evaluation} chains the execution, solving first one of the triple pattern and then restricting the second pattern to the results of the first.
  \item \emph{Interactive evaluation} is a more complex operation, in which
  two \ktrees are traversed simultaneously. The basic elements of this
  strategy include a synchronized traversal of the conceptual trees. Regardless of the type or complexity of the join operation, the essential steps
  of interactive evaluation are based on the access to one or more nodes in the conceptual trees
  of different \ktrees, operations that are directly supported by \dktrees.
\end{itemize}

\subsubsection{Update operations using \dktrees}

Our proposal is able to answer all the basic queries supported by
\ktriples simply replacing static \ktrees by \dktrees. Next, we will
show that update operations in RDF datasets can also be easily
supported by our proposal. This is presented as a proof of concept
of the applicability of \dktrees to this domain, even though our
proposal focuses only on the representation of the triples.

The most usual update operation in an RDF dataset is probably the
insertion of new triples, either one by one or, more frequently, in
small collections corresponding to new information retrieved or
indexed regularly. The insertion of new triples in an existing RDF
database involves several operations in our representation based on
dictionary encoding:
\begin{itemize}
  \item First, the values of the subject, predicate and object of
  the new triple must be searched in the dictionary, and added
  if necessary. If all the elements existed in the dictionary
  the new triple is stored as a new entry in the \dktree corresponding to its predicate.
  \item If the triple corresponds to a new predicate, a
  new empty \dktree can be created to store the new subject-object pair.
  \item If the subject and/or object are new, we must add
  a new row/column to all the \dktrees. As explained before, this operation is usually
  trivial in \dktrees. In the worst case, we must increase the
  size of the matrices, an operation (with a cost comparable to the insertion
  of a new 1 in the matrix) that must be repeated in all the \dktrees.
\end{itemize}

The removal of triples to correct or delete wrong information is a
typical update operation in RDF datasets, as well. The possible
changes when triples are removed are similar to the insertion case:
when triples are removed we may need to simply remove a 1 from a
\dktree or remove a row/column (marking it as unused) if the
subject/objects has no associated triples.

We assume in our representation that subject-object elements are
stored together in the top-left region of the matrices, so that join
operations do not need to perform additional computations. This
allows us to focus on triple pattern queries and ignore the effect
of an RDF vocabulary. However, the insertion of triples may cause a
subject (object) to become a subject-object, and the deletion of
triples may transform a subject-object in just a subject or object.
A simple solution to avoid the problem in the dynamic case would be
to use a different setup where rows/columns of the matrices would
contain all the elements (subjects and objects) instead of storing
only subjects in the rows and objects in the columns. It should have
small effect in the overall compression, since \ktrees and \dktrees
depend mostly on the number and distribution of the 1s in the matrix
than on the matrix size.

In order to follow the original setup with subject-object elements
in the first rows/columns of the matrix, when a subject (object)
becomes a subject-object we need to move it to the \emph{beginning}
of the matrix. This requires finding all the 1s in the corresponding
row (column), allocating a new row and column at the beginning of
the matrix and inserting the 1s in the same locations in the new row
(column). Note that, even though we only described the ability of
\dktrees to add new rows at the end of the matrix, the process can
be trivially extended to add rows at the beginning: given an $n
\times n$ matrix, let us assume we place elements starting from the
center instead of doing it from the first row/column. With this
setup, we can add new subject-object elements from the center
towards the top-left corner (thus, expanding the matrix towards the
top-left corner), and append new subjects and objects in the usual
way (towards the bottom-right corner). That is, we can expand our
virtual matrix to add subjects and objects as usual, but also attach
new subject-object elements in unused rows/columns in the upper-left
section of the matrix. This change has small effect on compression
(the elements are still grouped essentially in the same way) and
allows us to keep the list of subject-object elements together,
hence following the same ordering of the static representation.


\subsection{Experimental evaluation on RDF datasets}

We compare our proposal based on \dktrees with the static data
structures used in \ktriples and its enhanced version,
\ktriplesplus, presented in \cite{Alvarez15}. The goal of these
experiments is to demonstrate the efficiency of \dktrees in this
context, and their ability to act as the basis for a dynamic compact
representation of RDF databases. Note that in \cite{Alvarez15}
\ktriples were already proved to be competitive with
state-of-the-art representations, both in compression and query
times. We will compare the space results and query times of dynamic
and static representations to show that \dktrees can store RDF
datasets with a reduced overhead over the space and time
requirements of a static representation.

\subsubsection{Experimental setup}

We experimentally compare our dynamic representation with the
equivalent one using static \ktrees. We use a collection of RDF
datasets of very different sizes and number of triples, and also
include datasets with few and many different predicates\footnote{The
datasets and general experimental setup used are based on the
experimental evaluation in \cite{Alvarez15}, where \ktriples and
\ktriplesplus are tested. We use the same datasets and query sets in
our tests.}. Table~\ref{tab:dktree:rdf:datasets} shows a summary
with some information about the datasets used. The dataset
\emph{jamendo}\footnote{\texttt{http://dbtune.org/jamendo}} stores
information about Creative Commons licensed music;
\emph{dblp}\footnote{\texttt{http://dblp.l3s.de/dblp++.php}} stores
information about computer science publications;
\emph{geonames}\footnote{\texttt{http://download.geonames.org/all-geonames-rdf.zip}}
stores geographic information; finally,
\emph{dbpedia}\footnote{\texttt{http://wiki.dbpedia.org/Downloads351}}
is a large dataset that extracts structured information from
Wikipedia. As shown in Table~\ref{tab:dktree:rdf:datasets}, the
number of predicates is small in all datasets except \emph{dbpedia},
that is also the largest dataset and will be the best example to
measure the scalability of queries with variable predicate.

\begin{table}[ht!]
\centering

\begin{tabular}{|c|r|r|r|r|}
\hline
Collection  & \#triples         & \#predicates  &   \#subjects  &   \#objects   \\
\hline
jamendo     &   1,049,639       & 28            &   335,926     &   440,604     \\
dblp        &   46,597,620      & 27            &   2,840,639   &   19,639,731  \\
geonames    &   112,235,492     & 26            &   8,147,136   &   41,111,569  \\
dbpedia     &   232,542,405     & 39,672        &   18,425,128  &   65,200,769  \\
\hline
\end{tabular}
\caption{RDF datasets used in our experiments.}
\label{tab:dktree:rdf:datasets}
\end{table}

We build our dynamic representation following the same
procedure used for static \ktrees, and a similar setup. Elements that are both
subject and object in any triple pattern are grouped in the first
rows/columns of the matrices. We use a hybrid \ktree
representation, with $\k=4$ in the first 5 levels of decomposition
and $\k=2$ in the remaining levels. The \dktree uses a sampling
factor $s=128$ in the leaf blocks, while
static \ktrees use a single-level rank implementation that samples
every 20 integers (80 bytes). In \dktrees we use a block size
$B=512$ and $e+1=4$ different block sizes.

We test query times in all the approaches for all possible triple
patterns (except $(?S,?P,?O)$, that simply retrieves the complete
dataset) and some join queries involving just 2 triple patterns. We
use the experimental testbed\footnote{The full testbed is available
at \texttt{http://dataweb.infor.uva.es/queries-k2triples.tgz}} in
\cite{Alvarez15} to directly compare our representation with
\ktriples. To test triple patterns, we use a query set including 500
random triple patterns for each dataset and pattern. To test join
operations we use query sets including 50 different queries,
randomly selected from a larger group of 500 random queries and
divided in two groups: 25 have a number of results above the average
and 25 have a number of results below the average.

The experimental evaluation of join operations is expected to yield
similar comparison results to triple patterns, considering the fact
that the implementation of the different strategies to answer join
queries is identical in \dktrees and static \ktrees. As a proof of
concept of the applicability of \dktrees in more complex queries, we
will experimentally evaluate \dktrees, and \ktriples to answer join
operations $(?V,P_1,O_1)\bowtie(?V,P_2,O_2)$ (\emph{join 1}: only
the join variable is undetermined) and
$(?V,P_1,O_1)\bowtie(?V,?P_2,O_2)$ (\emph{join 2}: one of the
predicates is variable). Notice that each join type can lead to 3
different join operations depending on whether the join variable is
subject or object in each of the triple patterns: for example, join
1 can be of the form $(?V,P_1,O_1)\bowtie(?V,P_2,O_2)$ (S-S),
$(S_1,P_1,?V)\bowtie(?V,P_2,O_2)$ (S-O) and
$(S_1,P_1,?V)\bowtie(S_2,P_2,?V)$ (O-O).

\subsubsection{Space results}

We compare the space requirements of our dynamic representation,
based on \dktrees, with \ktriples and its improvement,
\ktriplesplus, in all the studied datasets. We select the \ktree
representations that obtain the best compression results: static
\ktrees used in \ktriples and \ktriplesplus use a matrix vocabulary
with $\k'=8$; \dktrees do not use a matrix vocabulary.
Table~\ref{table:rdfCollections} shows the total space requirements
on the different collections studied.

\begin{table}[ht!]
\centering

{\fontsize{9}{11}\selectfont
\begin{tabular}{|c|r|r|r|r|r|}
\hline
Collection & \ktriples  & \ktriplesplus  &  \dktrees\\
\hline
jamendo     &   0.74    &   1.28        &   1.61        \\
dblp        &   82.48   &   99.24       &   125.34      \\
geonames    &   152.20  &   188.63      &   242.60      \\
dbpedia     &   931.44  &   1,178.38    &   1,151.90    \\
\hline
\end{tabular}
} \caption{Space results for all RDF collections (sizes in MB).}
\label{table:rdfCollections}

\end{table}

Our dynamic representation is significantly larger than the
equivalent static version, \ktriples, in all the datasets. In
\emph{jamendo}, a very small dataset,  the dynamic representation
requires more than twice the space of \ktriples. However, the
overhead required by the dynamic version is smaller in larger
datasets and particularly in \emph{dbpedia}. The \dktree with no matrix
vocabulary is also able to store the dataset with an overhead below 50\%
extra in the \emph{dblp} and \emph{geonames} datasets. Even though
the overhead is significant, the results are still relevant since \ktriples was proved to be several times smaller
than other RDF stores like MonetDB and RDF-3X in these datasets (at
least 4 times smaller than MonetDB, the second-best approach in
space, in all the datasets except \emph{dbpedia}~\cite{Alvarez15}).

In the \emph{dbpedia} dataset our proposal has a space overhead
around 20\% over \ktriples, and becomes smaller than the
\ktriplesplus static representation. This result is mostly due to
the characteristics of the \emph{dbpedia} dataset, that contains many
predicates with few triples. The static representations based
on \ktriples store a static \ktree representation for each different
predicate, each one containing its own matrix vocabulary. The utilization
of a matrix vocabulary does not improve
compression in these matrices. However, most of the cost of the representation is in
the matrices with many triples, so the matrix vocabulary still obtains the best results overall.

\subsubsection{Query times}

\paragraph*{Triple patterns}
We first measure the efficiency of our dynamic proposal in
comparison with \ktriples to answer simple queries (triple patterns)
in all the studied datasets. The results for all the datasets are
shown in different tables: Table \ref{tab:dk2tree:rdfTimesJAMENDO}
shows the results for \emph{jamendo}; Table
\ref{tab:dk2tree:rdfTimesDBLP}, the results for \emph{dblp}; Table
\ref{tab:dk2tree:rdfTimesGEONAMES}, for \emph{geonames} and Table
\ref{tab:dk2tree:rdfTimesDBPEDIA}, for \emph{dbpedia}. For each
dataset we show the query times of \ktriples, \ktriplesplus (only in
queries with variable predicate) and our equivalent dynamic
representation of \ktriples. The last row of each table shows the
ratio between our dynamic representation and \ktriples, as an
estimation of the relative efficiency of \dktrees.

\begin{table}[ht!]
{\footnotesize
    \centering
    \begin{tabular}{|l|r r r r| r r r |}
        \hline
         & $S,P,O$ & $S,P,?O$ & $?S,P, O$ & $?S,P,?O$ & $S,?P,O$ & $S,?P,?O$ & $?S,?P,O$ \\
        \hline
        \ktriples & 1.0 & 4.6 & 102.8 & 6954.1 & 4.9 & 39.4 & 29.3 \\
        \ktriplesplus &  &  &  &  & 1.1 & 23.6 & 10.0 \\
        Dynamic & 1.9 & 4.8 & 235.6 & 12788.5 & 6.0 & 34.6 & 28.4 \\
        \hline
        Ratio & 1.88 & 1.06 & 2.29 & 1.84 & 1.22 & 0.88 & 0.97 \\
        \hline
    \end{tabular}
} \caption{Query times for triple patterns in \emph{jamendo}. Times
in $\mu$s/query.} \label{tab:dk2tree:rdfTimesJAMENDO}

\end{table}

\begin{table}[ht!]
{\footnotesize \centering
\begin{tabular}{|l|r r r r| r r r |}
\hline
Solution & $S,P,O$ & $S,P,?O$ & $?S,P, O$ & $?S,P,?O$ & $S,?P,O$ & $S,?P,?O$ & $?S,?P,O$ \\
\hline
\ktriples & 1.2 & 79.8 & 1016.4 & 771061.6 & 3.6 & 1294.1 & 187.5 \\
\ktriplesplus &  &  &  &  & 1.7 & 1102.1 & 140.8 \\
Dynamic & 6.5 & 92.9 & 2776.3 & 1450058.3 & 14.0 & 1421.8 & 247.9 \\
\hline
Ratio & 5.54 & 1.16 & 2.73 & 1.88 & 3.87 & 1.10 & 1.32 \\
\hline
\end{tabular}
} \caption{Query times for triple patterns in \emph{dblp}. Times in
$\mu$s/query.} \label{tab:dk2tree:rdfTimesDBLP}

\end{table}

\begin{table}[ht!]
{\footnotesize \centering
\begin{tabular}{|l|r r r r| r r r |}
\hline
Solution & $S,P,O$ & $S,P,?O$ & $?S,P, O$ & $?S,P,?O$ & $S,?P,O$ & $S,?P,?O$ & $?S,?P,O$ \\
\hline
\ktriples & 1.2 & 59.4 & 4588.0 & 1603677.4 & 2.9 & 1192.9 & 273.7 \\
\ktriplesplus &  &  &  &  & 1.4 & 915.9 & 139.0 \\
Dynamic & 9.4 & 79.6 & 9544.2 & 2958262.4 & 17.9 & 1514.7 & 423.5 \\
\hline
Ratio & 7.71 & 1.34 & 2.08 & 1.84 & 6.11 & 1.27 & 1.55 \\
\hline
\end{tabular}
} \caption{Query times for triple patterns in \emph{geonames}. Times
in $\mu$s/query.} \label{tab:dk2tree:rdfTimesGEONAMES}

\end{table}

\begin{table}[ht!]
{\footnotesize \centering
\begin{tabular}{|l|r r r r| r r r |}
\hline
Solution & $S,P,O$ & $S,P,?O$ & $?S,P, O$ & $?S,P,?O$ & $S,?P,O$ & $S,?P,?O$ & $?S,?P,O$ \\
\hline
\ktriples & 1.1 & 441.4 & 10.5 & 1859.5 & 7960.3 & 54497.4 & 29447.7 \\
\ktriplesplus &  &  &  &  & 1.4 & 2216.7 & 518.3 \\
Dynamic & 6.6 & 561.9 & 19.1 & 3870.7 & 23045.4 & 83340.2 & 57051.8 \\
\hline
Ratio & 6.21 & 1.27 & 1.82 & 2.08 & 2.90 & 1.53 & 1.94 \\
\hline
\end{tabular}
} \caption{Query times for triple patterns in \emph{dbpedia}. Times
in $\mu$s/query.} \label{tab:dk2tree:rdfTimesDBPEDIA}

\end{table}

In most of the datasets and queries, query times of \dktrees are
between 1.2 and 2 times higher than in \ktriples. The results in
Table \ref{tab:dk2tree:rdfTimesJAMENDO} for the dataset
\emph{jamendo} show some anomalies, with the \dktree performing
faster than a static representation. However, due to the reduced
size of the dataset we shall disregard these results and focus on
the larger datasets. The results in Table
\ref{tab:dk2tree:rdfTimesDBLP}, Table
\ref{tab:dk2tree:rdfTimesGEONAMES} and Table
\ref{tab:dk2tree:rdfTimesDBPEDIA} show very different query times
but the ratios shown in each table are very similar in all three
datasets.

Our results evidence that \dktrees are several times slower than
static \ktrees in triple patterns that are implemented with
single-cell retrieval queries (i.e. patterns $(S,P,O)$ and
$(?S,P,?O)$). Particularly, \dktrees are 5.5-7.7 times slower than
static \ktrees to answer $(S,P,O)$ queries. In this queries, the
cost of accessing \ttree and \ltree is very high since a single
position is accessed per level of the tree.

In all the remaining patterns, i.e. those that are translated into row/column
or full-range queries in one or many \ktrees, \dktrees are much more
competitive with static \ktrees, obtaining query times less than 2
times slower than \ktriples in most cases. These differences are
mostly due to the indexed representation used in \dktrees, that
avoids complete traversals of \ttree or \ltree when many close
positions are accessed in each query. In all these patterns,
multiple positions are accessed at each level of the \dktree, in
many cases these positions are actually in the same leaf node of \ttree or \ltree,
so the additional cost of traversing \ttree or \ltree is greatly diminished.

Let us focus now on the effect of the additional indexes used in
\ktriplesplus. In the datasets with few predicates, that are the
most usual RDF datasets, \ktriplesplus is around 1.5-3 times faster
than \ktriples, hence up to 10 times faster than \dktrees in
$(S,?P,O)$ queries and up to 3 times faster in the remaining
patterns with variable predicate. In the \emph{dbpedia} dataset, the relative
efficiency of \ktriplesplus is much more significant, reducing query
times by several orders of magnitude. This makes \dktrees much
slower than \ktriplesplus in this dataset to answer patterns with
variable predicate.

\paragraph*{Join operations}

Next we test the efficiency of \dktrees in comparison with \ktriples
to answer join queries. Considering the results obtained with single triple patterns,
and in order to better show the relative efficiency of our dynamic representation,
we focus on the join operations that are more selective, or require more selective
operations in the matrices.

We start our experiments with \emph{join 1}
($(?V,P_1,O_1)\bowtie(?V,P_2,O_2)$). We test the 3 different join
strategies applied to this join: \emph{i)} independent evaluation
requires two row/column queries in different \ktrees and an
intersection of the results; \emph{ii)} in chain evaluation, we run
a row/column query in the \ktree corresponding to $P_1$ and for each
result $v$ obtained we run a single cell retrieval in the \ktree for
$P_2$; \emph{iii)} the interactive evaluation runs two synchronized
row/column queries in both \ktrees. We test all the join categories
S-O, S-S and O-O and query sets with few results (small) or more
results (big).

\begin{figure}[ht!]
\begin{center}
   \includegraphics[width=\textwidth]{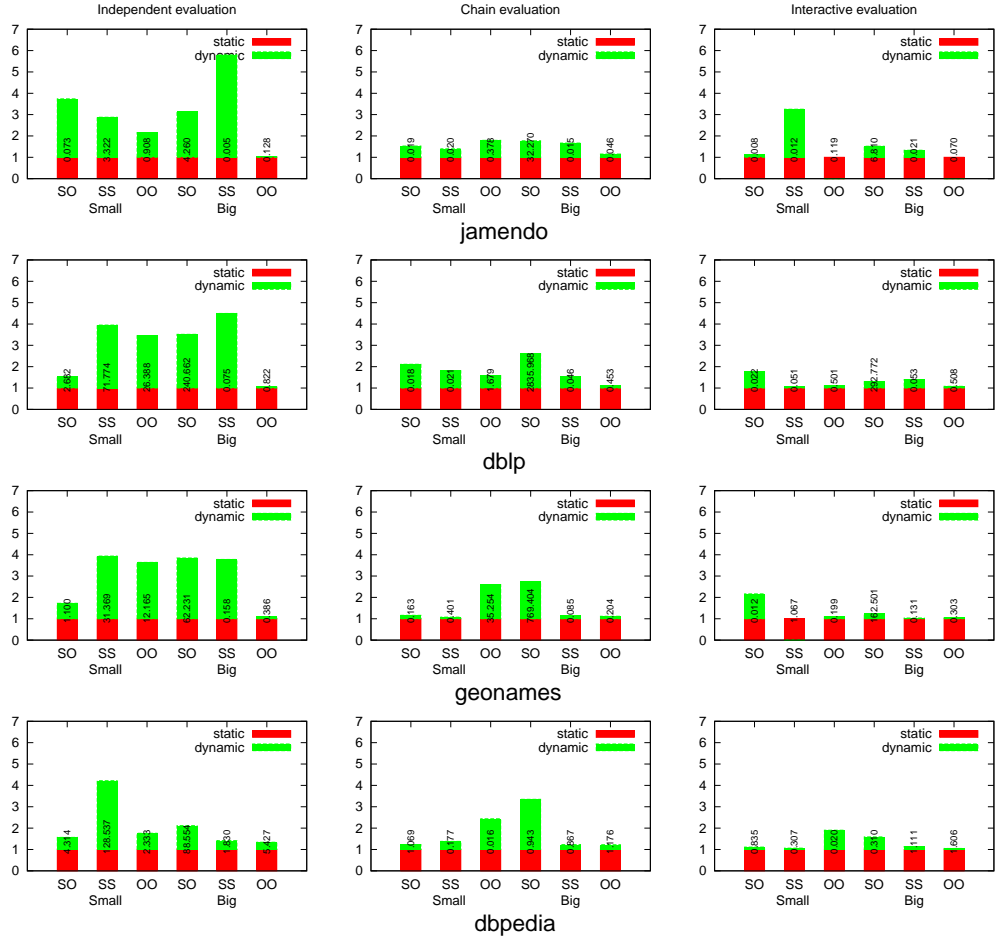}
   \caption{Comparison of static and dynamic query times in \emph{join 1} in all the studied datasets and query types. Results are normalized by static query times. Query times of the static representation are shown in ms/query.}
    \label{fig:dk2tree:join1}
\end{center}
\end{figure}

Figure~\ref{fig:dk2tree:join1} shows the results for \emph{join 1}.
Given the significant differences in query times between datasets,
strategies and even query sets, we normalize all the results so that
the query times of static \ktrees are always at the same level. The
height of the bar for the static representation will always be 1,
and the bar for the \dktree representation shows the actual overhead
of our dynamic representation. The actual query times in the static
representation are also displayed, in ms/query.

Results show that \dktrees are very competitive with \ktriples in
most of the datasets and strategies: independent evaluation (left
plots of Figure~\ref{fig:dk2tree:join1}) yields the worst results
for \dktrees, that are 3-4 times slower than \ktriples in most of
the query sets, except on the larger dataset \emph{dbpedia} where
our solution is on average less than 2 times slower than static
\ktrees. In chain evaluation \dktrees are less than 2 times slower
than a static representation in most of the datasets and query sets.
Finally, if we use the interactive evaluation strategy \dktrees are
able to obtain query times very close to static \ktrees in most of
the cases.

Averaging the results in all the datasets and query sets for each
evaluation strategy, \dktrees are 3.5 times slower than static
\ktrees in the independent evaluation strategy, 2.6 times slower
using the chain evaluation strategy and only 27\% slower using the interactive evaluation strategy.
Even though the results vary significantly between datasets and query sets, the
overall results show that \dktrees are able to support the query
operation with a reasonable overhead in space and query times.

\begin{figure}[ht!]
\begin{center}
   \includegraphics[width=\textwidth]{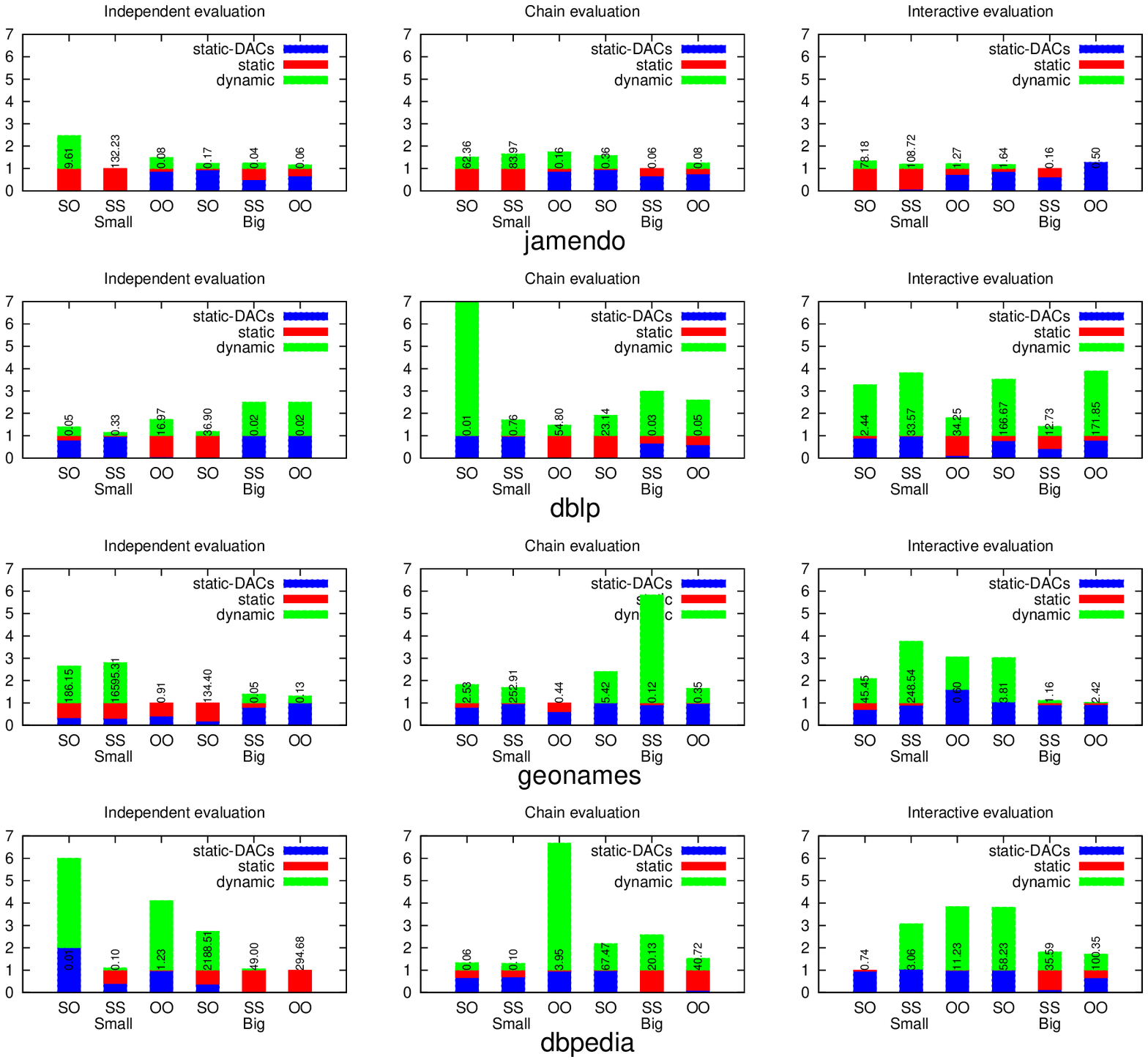}
   \caption{Comparison of static and dynamic query times in \emph{join 2} in all the studied datasets and query types. Results are normalized by query times of \ktriples. Query times of the static representation are shown in ms/query.}
    \label{fig:dk2tree:join2}
\end{center}
\end{figure}

\paragraph*{Join queries with variable predicate}

Next we analyze our proposal in join operations with variable
predicates. We compare \dktrees with static \ktriples and
\ktriplesplus (static-DACs) to answer the \emph{join 2}
($(?V,P_1,O_1)\bowtie(?V,?P_2,O_2)$), that involves a variable
predicate.

The results are shown in Figure \ref{fig:dk2tree:join2}. Bar heights are also normalized
by the static query times in these graphs. Again, we obtain diverse results in the comparison depending
on the dataset, the evaluation strategy and the query set. In spite
of the varying results, \dktrees show again an overhead in query time limited by a small factor:
our dynamic representation is between 1.5 and 3 times slower than \ktriples.

If we compare \dktrees with the best static representation, that is
now \ktriplesplus, the overhead required by the dynamic
representation becomes larger. Notice that in
Figure~\ref{fig:dk2tree:join2} there are some query sets for which
the query times of \ktriplesplus (\emph{static-DACs}) are so much
lower than \ktriples that the result for \ktriplesplus is not even
visible in the plot. This is consistent with the results obtained in
triple patterns, where the use of S-P and O-P indexes in
\ktriplesplus resulted in a major improvement. However, in many
cases the improvement obtained by \ktriplesplus is not so
significant and \dktrees are still reasonably close in query times.

\section{Conclusions} \label{sec:conclusions}

The compact representation of static binary relations can be useful in many contexts,
but in many application areas relations may be subject to frequent changes.
We have introduced the \dktree, a representation of binary relations with support for update operations.
The \dktree is an evolution of the \ktree, an inherently static data structure that
obtained great compression and query times in different types of binary relations.
Our dynamic representation is able to obtain good compression and query times, relatively close
to those of static \ktrees, and provides at the same time support for the most usual update operations:
insertion and deletion of pairs in the binary relation, and also changes in the base sets A and B of the relation.

Our experimental evaluation shows that the \dktree, even though it
has a significant overhead compared with the static representation,
is able to obtain results close enough to those of the static
representation to be competitive in areas where the original static
representation was already efficient. In particular, we demonstrate
that our representation obtains compression results close to a
static data structure in Web graphs, and we study the overhead
required in the representation of RDF graphs. Our representation
performs in general at the same order of magnitude than static
\ktrees in this domain, where state-of-the-art alternatives are
orders of magnitude larger or slower. Additionally, in many
operations and datasets the overhead required by our representation
becomes small enough (around 20\% space or time overhead in
different cases) to be a good representation even in a context of
dynamic binary relations with low rate of changes.


\end{document}